%% file: DDALPHA.tex
\documentclass[12pt,a4paper]{article}
\pdfoutput=1

\newcommand\argmax{\operatornamewithlimits{argmax}}
\newcommand\argmin{\operatornamewithlimits{argmin}}

\newenvironment{algenum}{
\vspace{-\topsep}
\begin{enumerate}
  \setlength{\itemsep}{0pt}
  \setlength{\parskip}{0pt}
  \setlength{\parsep}{0pt}
}{\end{enumerate}\vspace{-\topsep}}

\usepackage{comment}
\usepackage{longtable}
\usepackage{multirow}
\usepackage{multicol}
\usepackage{dcolumn}
\newcolumntype{d}[1]{D{.}{.}{#1}}
\newcolumntype{d}{D{.}{.}{2.3}}

\def\LBL{\textsf{\textbf{---}}} 
\def\LBD{{\textbf{- -}}} 

\usepackage{amsmath}
\usepackage{times}
\usepackage{bm}
\usepackage{natbib}
\usepackage{amssymb}
\usepackage{graphicx}
\usepackage{color}
\usepackage{verbatim}
\usepackage[colorlinks,citecolor=blue,urlcolor=blue,linkcolor=blue]{hyperref}

\makeatletter
\define@key{Gin}{Trim}
           {\let\Gin@viewport@code\Gin@trim\expandafter\Gread@parse@vp#1 \\}
\makeatother

\usepackage{lscape}

\usepackage{overpic}

\newcommand\tab{\phantom{tabs}}

\newcommand{\trimval}{}

\usepackage{calc}
\newlength{\textwnew}
\usepackage[width=1.085\textwidth,font=small]{caption}

\newenvironment{Code}{
\verbatim
}{\endverbatim
}

\usepackage{listings}
\newcommand{\code}[1]{\lstinline|#1|}

\let\proglang=\textsf
\newcommand{\pkg}[1]{{\fontseries{b}\selectfont #1}}

\begin{document}

\raggedbottom

\title{Depth and Depth-Based Classification with \proglang{R}-Package \pkg{ddalpha}}
\author{Oleksii Pokotylo\\{\small University of Cologne} \and
        Pavlo Mozharovskyi\\{\small Agrocampus Ouest} \and
        Rainer Dyckerhoff\\{\small University of Cologne}
        }
\date{August 14, 2016}
\maketitle

\abstract{
  Following the seminal idea of \cite{Tukey75}, data depth is a function that  measures how close an arbitrary point of the space is located to an implicitly defined center of a data cloud. Having undergone theoretical and computational developments, it is now employed in numerous applications with classification being the most popular one. The \proglang{R}-package \pkg{ddalpha} is a software directed to fuse experience of the applicant with recent achievements in the area of data depth and depth-based classification.

  \pkg{ddalpha} provides an implementation for exact and approximate computation of most reasonable and widely applied notions of data depth. These can be further used in the depth-based multivariate and functional classifiers implemented in the package, where the $DD\alpha$-procedure is in the main focus. The package is expandable with user-defined custom depth methods and separators. The implemented functions for depth visualization and the built-in benchmark procedures may also serve to provide insights into the geometry of the data and the quality of pattern recognition.
}
\indent\\


\sloppy

\section[Introduction]{Introduction}\label{sec:intro}

In 1975 John W. \citeauthor{Tukey75}, in his work on mathematics and the picturing of data, proposed a novel way of data description, which evolved into a measure of multivariate centrality named \emph{data depth}. For a data sample, this statistical function determines centrality, or representativeness of an arbitrary point in the data, and thus allows for multivariate ordering of data regarding their centrality. More formally, given a data cloud $\boldsymbol{X}=\{\boldsymbol{x}_1,...,\boldsymbol{x}_n\}$ in $\mathbb{R}^d$, for a point $\boldsymbol{z}$ of the same space, a depth function $D(\boldsymbol{z}|\boldsymbol{X})$ measures how \emph{close} $\boldsymbol{z}$ is located to some (implicitly defined) \emph{center} of $\boldsymbol{X}$. Different concepts of closeness between a point $\boldsymbol{z}$ and a data cloud $\boldsymbol{X}$ suggest a diversity of possibilities to define such a function and a center as its maximizer. Naturally, each depth notion concentrates on a certain aspect of $\boldsymbol{X}$, and thus possesses various theoretical and computational properties.
Many depth notions have arisen during the last several decades
differing in properties and being suitable for various applications.
Mahalanobis \citep{Mahalanobis36}, halfspace \citep{Tukey75}, simplicial volume \citep{Oja83}, simplicial \citep{Liu90}, zonoid \citep{KoshevoyM97}, projection \citep{ZuoS00}, spatial \citep{VardiZ00} depths can be seen as well developed and most widely employed notions of depth function; see \cite{Mosler13} for a recent survey with details on categorization and properties.

Being intrinsically nonparametric, a depth function captures the geometrical features of given data in an affine-invariant way. By that, it appears to be useful for description of data's location, scatter, and shape, allowing for multivariate inference, detection of outliers, ordering of multivariate distributions, and in particular classification, that recently became an important and rapidly developing application of the depth machinery.
While the parameter-free nature of data depth ensures attractive theoretical properties of classifiers, its ability to reflect data topology provides promising predicting results on finite samples.

\subsection{Classification in the depth space}
Consider the following setting for supervised classification: Given a training sample consisting of $q$ classes $\boldsymbol{X}_1,...,\boldsymbol{X}_q$, each containing $n_i$, $i=1,...,q$, observations in $\mathbb{R}^d$. For a new observation $\boldsymbol{x}_0$, a class should be determined, to which it most probably belongs. Depth-based learning started with plug-in type classifiers. \citet{GhoshC05b} construct a depth-based classifier, which, in its na\"{i}ve form, assigns the observation $\boldsymbol{x}_0$ to the class in which it has maximal depth. 
They suggest an extension of the classifier, that is consistent w.r.t.~Bayes risk for classes stemming from elliptically symmetric distributions.
Further \citet{DuttaG11,DuttaG12} suggest a robust classifier and a classifier for $L_p$-symmetric distributions, see also \citet{CuiLY08}, \citet{MoslerH06}, and additionally \citet{Joernsten04} for unsupervised classification.

A novel way to perform depth-based classification has been suggested by \citet{LiCAL12}: first map a pair of training classes into a two-dimensional depth space, which is called the $DD$-plot, and then perform classification by selecting a polynomial that minimizes empirical risk. Finding such an optimal polynomial numerically is a very challenging and --- when done appropriately --- computationally involved task, with a solution that in practice can be unstable \citep[see][Section~1.2.2 for examples]{Mozharovskyi15}. In addition, 
the $DD$-plot should be rotated and the polynomial training phase should be done twice. Nevertheless, the scheme itself allows to construct optimal classifiers for wider classes of distributions than the elliptical family. Being further developed and applied by \cite{Vencalek11,LangeMM14a,MozharovskyiML15} it proved to be useful in practice, also in the functional setting \citep{MoslerM15,CuestaAlbertosFBOdlF15}.

The general depth-based supervised classification framework implemented in the \proglang{R}-package \pkg{ddalpha} can be described as follows. In the first training phase, each point of the training sample is mapped into the $q$-variate space of its depth values with respect to each of the classes $\boldsymbol{x}_i\mapsto\left(D(\boldsymbol{x}_i|\boldsymbol{X}_1),...,D(\boldsymbol{x}_i|\boldsymbol{X}_q)\right)$. In the second training phase, a low-dimensional classifier, flexible enough to account for the change in data topology due to the depth transform, is employed in the depth space. We suggest to use the $\alpha$-procedure, which is a nonparametric, robust, and computationally efficient separator. When classifying an unknown point $\boldsymbol{x}_0$, the first phase remains unchanged $\left(\boldsymbol{x}_0\mapsto\left(D(\boldsymbol{x}_0|\boldsymbol{X}_1),...,D(\boldsymbol{x}_0|\boldsymbol{X}_q)\right)\right)$, and in the second phase the trained $q$-variate separator assigns the depth-transformed point to one of the classes. Depth notions best reflecting data geometry share the common feature to attain value zero immediately beyond the convex hull of the data cloud. Thus, if such a data depth is used in the first phase, it may happen that $\boldsymbol{x}_0$ is mapped to the origin of the depth space, and thus cannot be readily classified. We call such a point an \emph{outsider} and suggest to apply a special treatment to assign it. If the data is of functional nature, a finitization step based on the \emph{location-slope \mbox{(LS-)} transform} precedes the above described process. Depth transform, $\alpha$-procedure, outsider treatment, and the preceding $LS$-transform constitute the $DD\alpha$-classifier. This together with the depth-calculating machinery constitutes the heart of the \proglang{R}-package \pkg{ddalpha}.

\subsection{The R-package ddalpha}
The \proglang{R}-package \pkg{ddalpha} is a software directed to fuse experience of the applicant with recent theoretical and computational achievements in the area of data depth and depth-based classification. It provides an implementation for exact and approximate computation of seven most reasonable and widely applied depth notions: Mahalanobis, halfspace, zonoid, projection, spatial, simplicial and simplicial volume depths. The variety of depth-calculating procedures includes functions for computation of data depth of one or more points w.r.t.~a data set, construction of the classification-ready $q$-dimensional depth space, visualization of the bivariate depth function for a sample in the form of upper-level contours and of a 3D-surface.

The main feature of the proposed methodology on the $DD$-plot is the $DD\alpha$-classifier, which is an adaptation of the $\alpha$-procedure to the depth space. Except for its efficient and fast implementation, \pkg{ddalpha} suggests other classification techniques that can be employed in the $DD$-plot: the original polynomial separator by \cite{LiCAL12} and the depth-based $k$NN-classifier proposed by \cite{Vencalek11}.

Halfspace, zonoid and simplicial depths
vanish beyond the convex hull of the sample, and thus cause outsiders during classification. For this case, \pkg{ddalpha} offers a number of outsider treatments and a mechanism for their management.

If it is decided to employ the $DD$-classifier, its constituents are to be chosen: data depth, classification technique in the depth space, and, if needed, outsider treatment and aggregation scheme for multi-class classification.
Their parameters, such as type and subset size of the variance-covariance estimator for Mahalanobis and spatial depth, number of approximating directions for halfspace and projection depth or part of simplices for approximating simplicial and simplicial volume depths, degree of polynomial extension for the $\alpha$-procedure or the polynomial classifier, number of nearest neighbors in the depth space or for an outsider treatment, \textit{etc.} must be set.
Rich built-in benchmark procedures allow to estimate the empirical risk and error rates of the $DD$-classifier and the portion of outsiders help in making the decision concerning the settings.

\pkg{ddalpha} possesses tools for immediate classification of functional data in which the measurements are first
brought onto a finite dimensional basis, and then fed to the depth-classifier. In addition, the componentwise classification technique by \cite{DelaigleHB12} is implemented.

Unlike other packages, \pkg{ddalpha} implements under one roof various depth functions and classifiers for multivariate and functional data.
\pkg{ddalpha} is the only package that implements zonoid depth and efficient exact halfspace depth. All depths in the package are implemented for any dimension $d\ge2$; except for the projection depth all implemented algorithms are exact, and supplemented by their approximating versions to deal with the increasing computational burden for large samples and higher dimensions. It also supports user-friendly definitions of depths and classifiers. 
In addition, the package contains 50 multivariate and 4 functional ready-to-use classification problems and data generators for a palette of distributions.

Most of the functions of the package are programmed in \proglang{C++}, in order to be fast and efficient.
The package has a module structure, which makes it expandable and allows user-defined custom depth methods and separators.
\pkg{ddalpha} employs \textbf{boost} (package \pkg{BH} \cite{BH}), a well known fast and widely applied library, and resorts to \pkg{Rcpp} \citep{Rcpp} allowing for calls of \proglang{R} functions from \proglang{C++}.

\newpage 
\subsection{Existing R-functionality on data depth}

Having proved to be useful in many areas, data depth and its applications find implementation in a number of \proglang{R}-packages: \pkg{aplpack}, \pkg{depth}, \pkg{localdepth}, \pkg{fda.usc}, \pkg{rsdepth}, \pkg{depthTools}, \pkg{MFHD}, \pkg{depth.plot}, \pkg{DepthProc}, \pkg{WMTregions}, \pkg{modQR}, \pkg{OjaNP}.

Functions of \pkg{aplpack} \citep{aplpack} allow to exactly compute bivariate halfspace depth and construct a bagplot.

\proglang{R}-package \pkg{depth} \citep{depth} provides implementation for the exact halfspace depth for $d\le 3$ and for approximate halfspace depth when $d\ge 3$, exact simplicial depth in $\mathbb{R}^2$, and exact simplicial volume depth in any dimension.

\pkg{DepthProc} \citep{DepthProc} calculates Mahalanobis, Euclidean, LP, bivariate regression, and modified band depth and approximates halfspace and projection depth, provides implementation for their local versions w.r.t.~\cite{PaindaveineVB13} and for the corresponding depth median estimators. It also contains functions for depth visualization and produces $DD$-plots.

\cite{AgostinelliR11} proposed local versions for several depth notions. Their \proglang{R}-package \pkg{localdepth} \citep{localdepth} evaluates simplicial, univariate halfspace, Mahalanobis, ellipsoid depths and their localization.

The package \pkg{fda.usc} \citep{fda.usc} provides methods for exploratory and descriptive analysis of functional data. It contains functions for functional data representation, functional outliers detection, functional regression models and analysis of variance model, functional supervised and unsupervised classification. The package calculates the following depth functions for multivariate data: simplicial depth in $\mathbb{R}^2$, halfspace depth (aproximate for $d>3$), Mahalanobis depth, approximate projection depth, likelihood depth; and for functional data: Fraiman and Muniz depth, h-modal depth, random Tukey depth, random projection depth, double random projection depth. In addition the package suggests a number of classifiers acting in the functional depth space, namely maximum depth, polynomial, logistic regression, LDA, QDA, $k$NN and KDA ones.

Several packages have specific purpose or implement only one depth. Among the others are:
\pkg{depthTools} \citep{depthTools}, implementing different statistical tools for the description and analysis of gene expression data based on the modified band depth; 
\pkg{depth.plot} \citep{depth.plot}, containing the implementations of spatial depth and spatial ranks and constructing corresponding $DD$-plots; 
\pkg{MFHD} \citep{MFHD}, calculating multivariate functional halfspace depth and median for bivarite functional data;
\pkg{rsdepth} \citep{rsdepth}, implementing the ray shooting depth and the corresponding median in $\mathbb{R}^2$;
\pkg{WMTregions} \citep{WMTregions}, computing weighted-mean trimmed regions with zonoid regions as a special case;
\pkg{modQR} \citep{modQR}, calculating multiple-output regression quantiles with halfspace trimmed regions as a special case;
and
\pkg{OjaNP} \citep{OjaNP}, which offers efficient computation of the multivariate Oja median and related statistics.


\subsection{Outline of the article}\label{ssec:introOutline}

To facilitate understanding and keep the presentation solid, the functionality of the \proglang{R}-package \pkg{ddalpha} is illustrated through the article on the same functional data set ``ECG Five Days'' from \cite{UCRArchive}, which is a long ECG time series constituting two classes. 
The data set originally contains 890 objects.
We took a subset consisting of 70 objects only (35 from each of the days) which best demonstrates the general and complete aspects of the proposed procedures (\textit{e.g.}, existence of outliers in its bivariate projection or necessity of three features in the $\alpha$-procedure).

\begin{figure}[!h]
  \begin{center}
    \includegraphics[keepaspectratio=true,width = .495\textwnew, page=1]{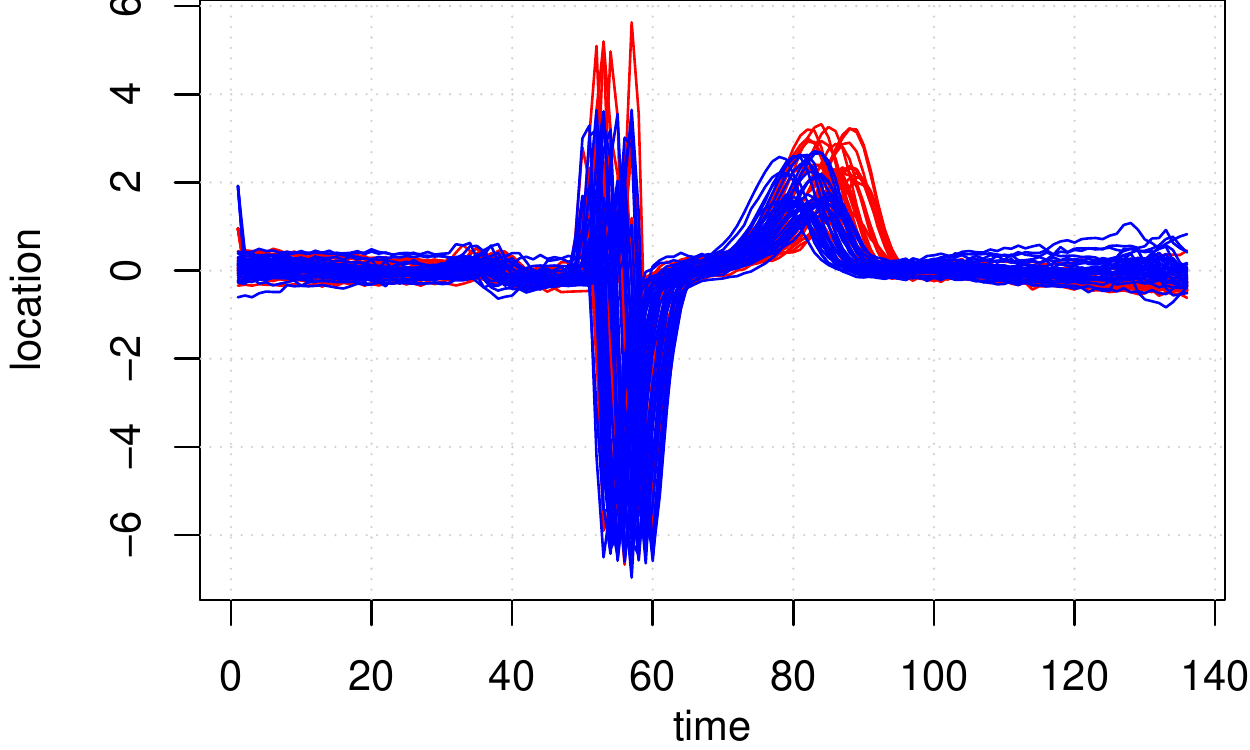}
    \includegraphics[keepaspectratio=true,width = .495\textwnew, page=2]{functional_data.pdf}
    \caption{ECG Five Days data (left) and their derivatives (right).}
    \label{fig:funcdata}
  \end{center}
\end{figure}

In Section~\ref{sec:functions} functional data are transformed into a finite dimensional space using the $LS$-transform, as it is shown in Figure~\ref{fig:funcdata} presenting the functions and their derivatives.
In this example we choose $L = S = 1$, and thus the $LS$-transform produces the two-dimensional discrete space, where each function is described by the area under the function and under its derivative as it is shown in Figure~\ref{fig:discretespace}, left.
Then in Section~\ref{sec:depth}, the depth is calculated in this two-dimensional space (Figure~\ref{fig:discretespace}, middle) and the $DD$-plot is constructed in Section~\ref{sec:ddclass} (Figure~\ref{fig:discretespace}, right).
The classification is performed by the $DD\alpha$-separator in the $DD$-plot. The steps of the $\alpha$-procedure are illustrated on Figure~\ref{tab:gt}.

Section~\ref{sec:depth} presents a theoretical description of the data depth and the depth notions implemented in the package. In addition, it compares their computation time and performance when employed in the maximum depth classifier.
Section~\ref{sec:ddclass} includes a comprehensive algorithmic description of the $DD\alpha$-classifier with a real-data illustration.
Further, it discusses other classification techniques that can be employed in the $DD$-plot.
The questions whether one should choose a depth that avoids outsiders or should allow for outsiders and classify them separately, and in which way, are considered in Section~\ref{sec:outsiders}.
Section~\ref{sec:functions} addresses the classification of functional data.
In Section~\ref{sec:application}, the basic structure and concepts of the \proglang{R}-package user interface are presented, along with a discussion of their usage for configuring the classifier and examples for calling its functions.

\section[Data depth]{Data depth}\label{sec:depth}

This section regards depth functions.
First (Section~\ref{ssec:depthConcept}), we briefly review the concept of data depth and its fundamental properties. 
Then (Section~\ref{ssec:depthNotions}), we give the definitions in their empirical versions
for several depth notions: Mahalanobis, projection, spatial, halfspace, simplicial, simplicial volume, zonoid depths.
For each notion, we shortly discuss relevant computational aspects, leaving motivations, ideas, and details to the corresponding literature and the software manual.
We do not touch the question of computation of depth-trimmed regions for the following reasons: first, for a number of depth notions there exist no algorithms; then, for some depth notions these can be computed using different \proglang{R}-packages, \textit{e.g.} \pkg{WMTregions} for the family of weighted-mean regions including zonoid depth \citep{BazovkinM12} or \pkg{modQR} for multiple-output quantile regression including halfspace depth as a particular case; finally, this is not required in classification.
After having introduced depth notions, we compare the speed of the implemented exact algorithms by means of simulated data (Section~\ref{ssec:depthComputation}).
The section is concluded (Section~\ref{ssec:depthMaxdepth}) by a comparison of error rates of the na\"{i}ve maximum depth classifier, paving a bridge to the more developed $DD$-plot classification which is covered in the following sections.

\subsection{The concept}\label{ssec:depthConcept}
Consider a point ${\boldsymbol z}\in\mathbb{R}^d$ and a data sample $\boldsymbol{X}=(\boldsymbol{x}_1,...,\boldsymbol{x}_n)^\prime$ 
in the $d$-dimensional Euclidean space, with $\boldsymbol{X}$ being a $(n\times d)$-matrix and $^\prime$ being the transposition operation.
A data depth is a function $D({\boldsymbol z}|\boldsymbol{X}):{\mathbb R}^d\mapsto[0,1]$ that describes how deep, or central, the observation ${\boldsymbol z}$ is located w.r.t.~$\boldsymbol{X}$. In a natural way, it involves some notion of center. This is any point of the space attaining the highest depth value in $\boldsymbol{X}$, and not necessarily a single one. In this view, depth can be seen as a center-outward ordering, \textit{i.e.}~points closer to the center have a higher depth, and those more outlying a smaller one.

The concept of a depth function can be formalized by stating postulates (requirements) it should satisfy. Following \citet{Dyckerhoff04} and \citet{Mosler13}, a \emph{depth function} is a function $D({\boldsymbol z}|\boldsymbol{X}):{\mathbb R}^d\mapsto[0,1]$ that is:
\begin{itemize}
    \item[(\emph{D1})] \emph{translation invariant}: $D({\boldsymbol z}+\boldsymbol b|\boldsymbol{X}+\boldsymbol{1}_n \boldsymbol b^\prime)=D({\boldsymbol z}|\boldsymbol{X})$ for all $\boldsymbol b\in{\mathbb R}^d$ (here $\boldsymbol{1}_n = (1,...,1)^\prime$),
    \item[(\emph{D2})] \emph{linear invariant}: $D(\boldsymbol A{\boldsymbol z}|\boldsymbol{X}\boldsymbol A^\prime)=D({\boldsymbol z}|\boldsymbol{X})$ for every nonsingular $d\times d$ matrix $\boldsymbol A$,
    \item[(\emph{D3})] \emph{zero at infinity}: $\lim_{\|{\boldsymbol z}\|\to\infty}D({\boldsymbol z}|\boldsymbol{X})=0$,
    \item[(\emph{D4})] \emph{monotone on rays}: Let ${\boldsymbol z}^*=\argmax_{{\boldsymbol z}\in{\mathbb R}^d}D({\boldsymbol z}|\boldsymbol{X})$, then for all $\,{\boldsymbol r}\in S^{d-1}$  the function $\beta\mapsto D({\boldsymbol z}^*+\beta{\boldsymbol r}|\boldsymbol{X})$ decreases in the weak sense, for $\,\beta>0$,
    \item[(\emph{D5})] \emph{upper semicontinuous}: the upper level sets $D_{\alpha}(\boldsymbol{X})=\{{\boldsymbol z}\in{\mathbb R}^d:D({\boldsymbol z}|\boldsymbol{X})\ge\alpha\}$ are closed for all $\,\alpha$.
\end{itemize}

For slightly different postulates see \cite{Liu92} and \cite{ZuoS00}.

The first two properties state that $D(\cdot|\boldsymbol{X})$ is \emph{affine invariant}. $\boldsymbol{A}$ in (D2) can be weakened to isometric linear transformations, which yields an \emph{orthogonal invariant} depth. Taking instead of $\boldsymbol{A}$ some constant $\lambda>0$ gives a \emph{scale invariant} depth function. (D3) ensures that the upper level sets $D_{\alpha}$, $\alpha>0$, are bounded. According to (D4), the upper level sets are starshaped around ${\boldsymbol z}^*$, and $D_{\max_{{\boldsymbol z}\in{\mathbb R}^d}D({\boldsymbol z}|\boldsymbol{X})}(\boldsymbol{X})$ is convex.
(D4) can be strengthened by requiring $D(\cdot|\boldsymbol{X})$ to be a \emph{quasiconcave} function. In this case, the upper level sets are convex for all $\,\alpha>0$. (D5) is a useful technical restriction.

Upper level sets $D_{\alpha}(\boldsymbol{X}) = \{\boldsymbol{x}\in\mathbb{R}^d:D(\boldsymbol{x}|\boldsymbol{X})\ge\alpha\}$
of a depth function are also called \emph{depth-trimmed} or \emph{central regions}. They describe the distribution's location, dispersion, and shape. For given $\boldsymbol{X}$, the sets $D_{\alpha}(\boldsymbol{X})$ constitute a nested family of trimming regions. Note that due to (D1) and (D2) the central regions are affine equivariant, due to (D3) bounded, due to (D5) closed, and due to (D4) star-shaped (respectively convex, if quasiconcaveness of $D(\cdot|\boldsymbol{X})$ is additionally required).

\subsection{Implemented notions}\label{ssec:depthNotions}
The \proglang{R}-package \pkg{ddalpha} implements a number of depths. Below we consider their empirical versions.
For each implemented notion of data depth, the depth surface (left) and depth contours (right) are plotted in Figures~\ref{fig:bivariate_depths1} and~\ref{fig:bivariate_depths2} for bivariate data used in Section~\ref{sec:functions}.

\def\wb{.80\textwnew}
\def\ww{.3\textwnew}
\def\ws{.05\textwnew}
\fboxsep0.1mm
\begin{figure}
     \begin{center}
\fbox{\vbox{\hsize=\wb\includegraphics[keepaspectratio=true,width = \ww, trim = 27mm 6.5mm 28mm 54mm, clip,scale=0.80]{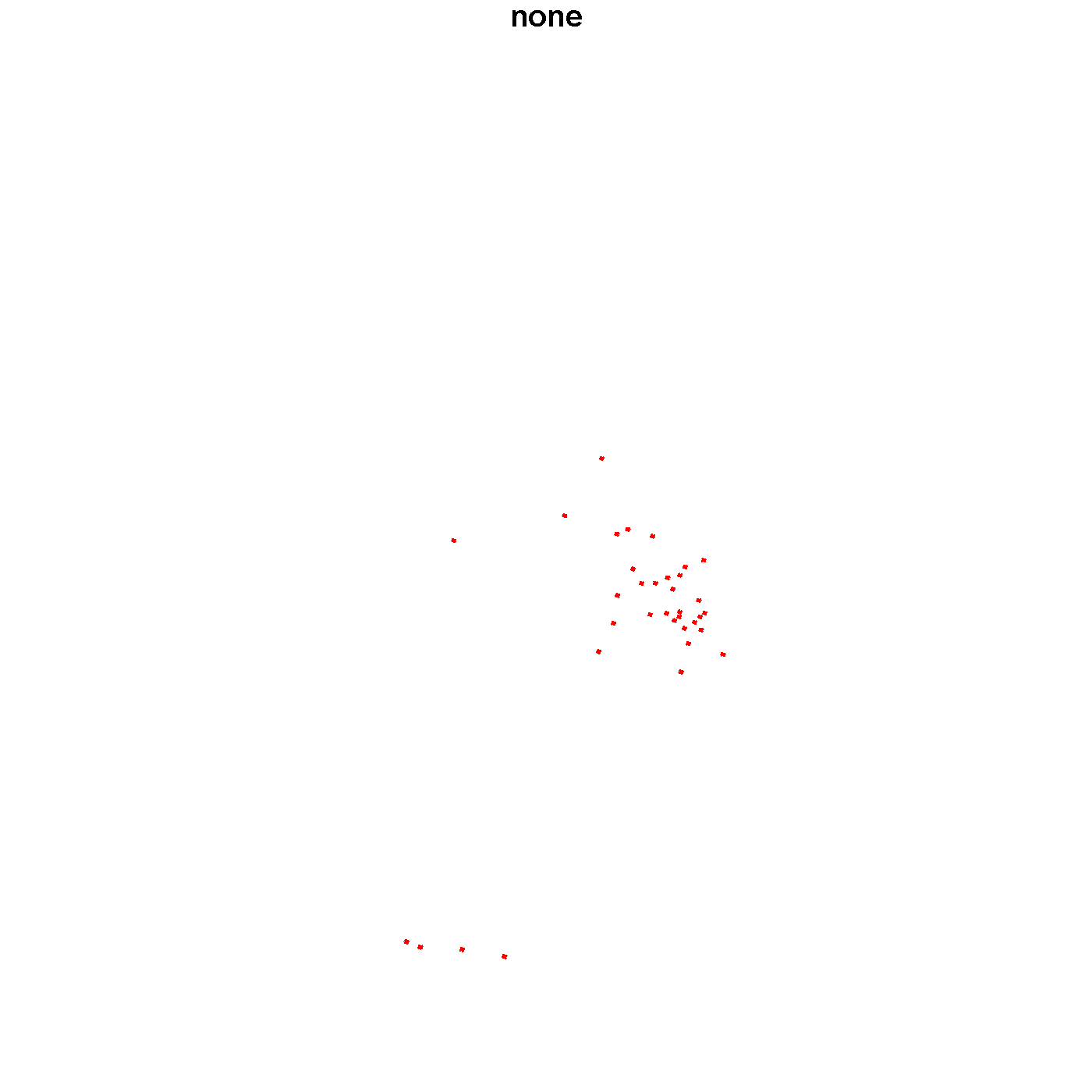}
\hspace{\ws}\includegraphics[keepaspectratio=true,width = \ww, trim = 10mm 16mm 10mm 20mm, clip, page = 1]{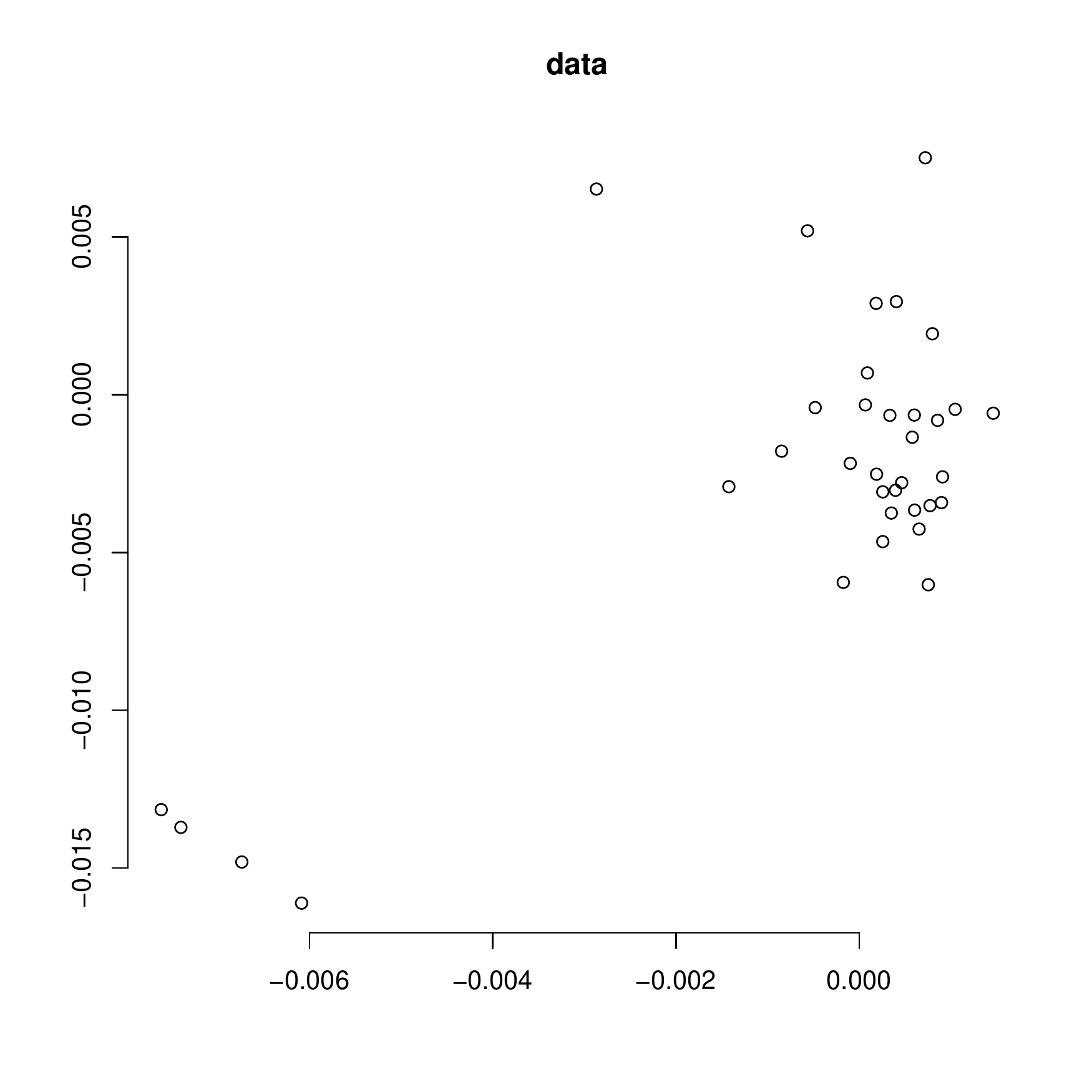}\\
{\small Bivariate data}}}
\fbox{\vbox{\hsize=\wb\includegraphics[keepaspectratio=true,width = \ww, trim = 27mm 6.5mm 28mm 54mm, clip,scale=0.80]{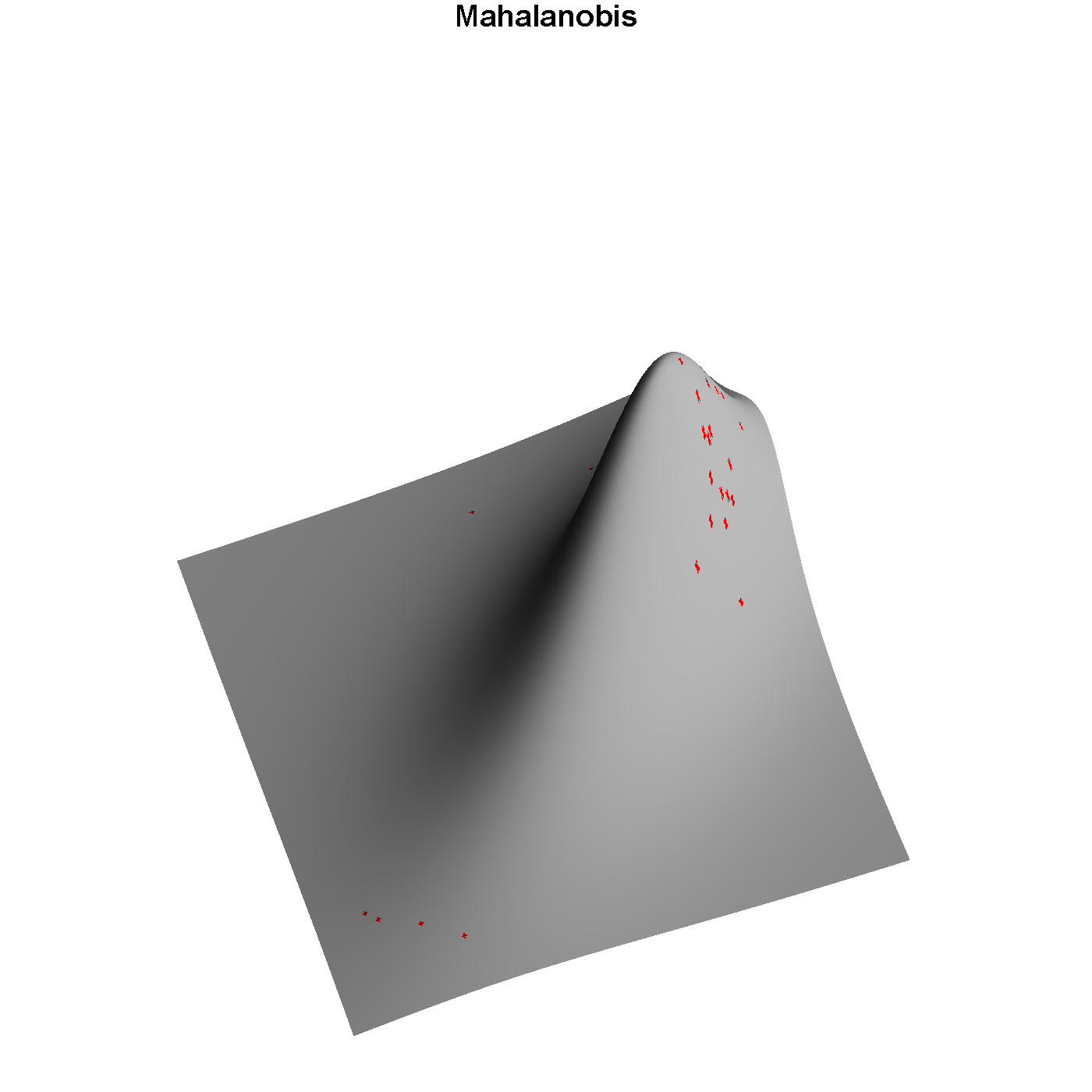}
\hspace{\ws}\includegraphics[keepaspectratio=true,width = \ww, trim = 10mm 16mm 10mm 20mm, clip, page = 3]{depth_contours_noborder.pdf}\\
{\small Mahalanobis depth}}}
\fbox{\vbox{\hsize=\wb\includegraphics[keepaspectratio=true,width = \ww, trim = 27mm 6.5mm 28mm 54mm, clip,scale=0.80]{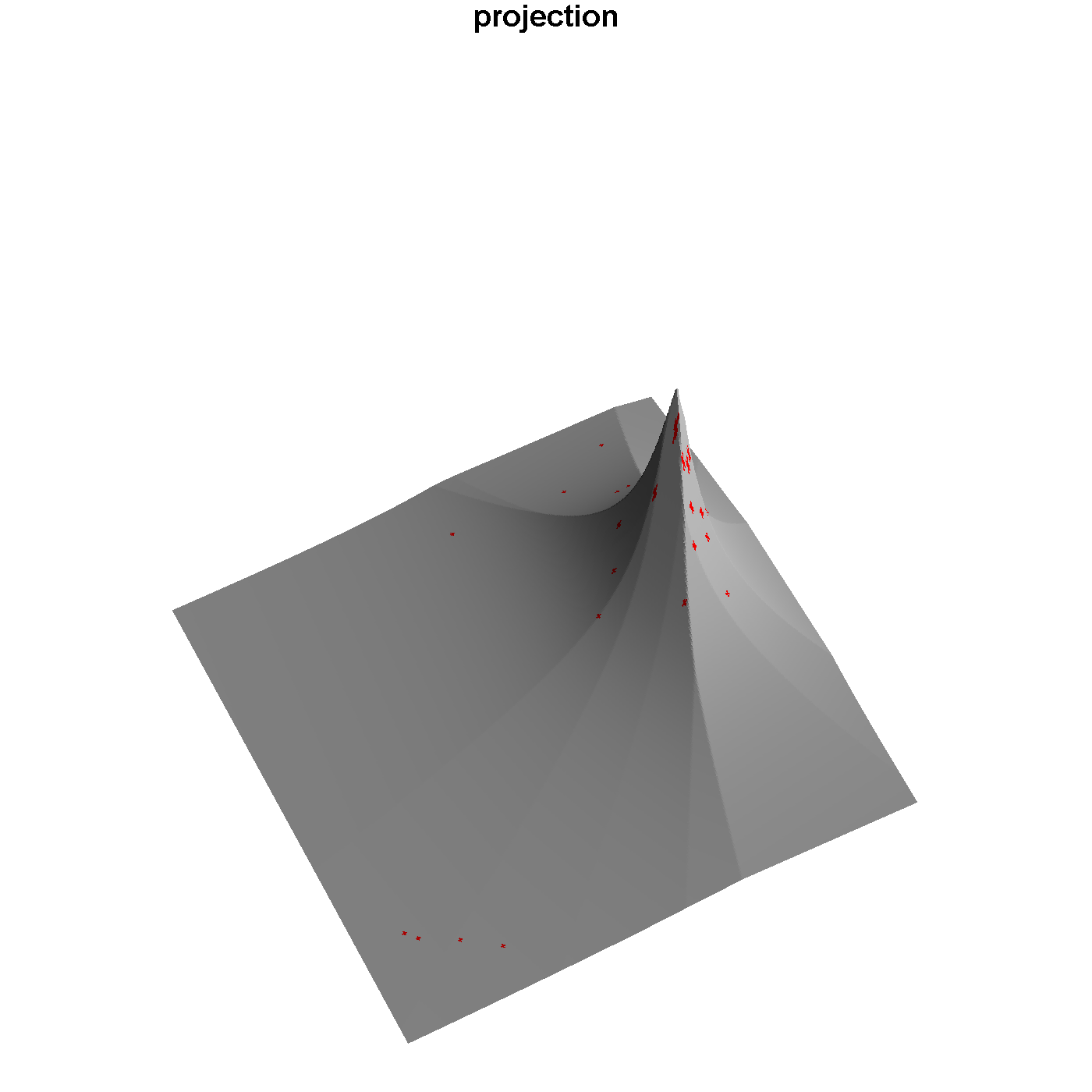}
\hspace{\ws}\includegraphics[keepaspectratio=true,width = \ww, trim = 10mm 16mm 10mm 20mm, clip, page = 4]{depth_contours_noborder.pdf}\\
{\small Projection depth}}}
\fbox{\vbox{\hsize=\wb\includegraphics[keepaspectratio=true,width = \ww, trim = 27mm 6.5mm 28mm 54mm, clip,scale=0.80]{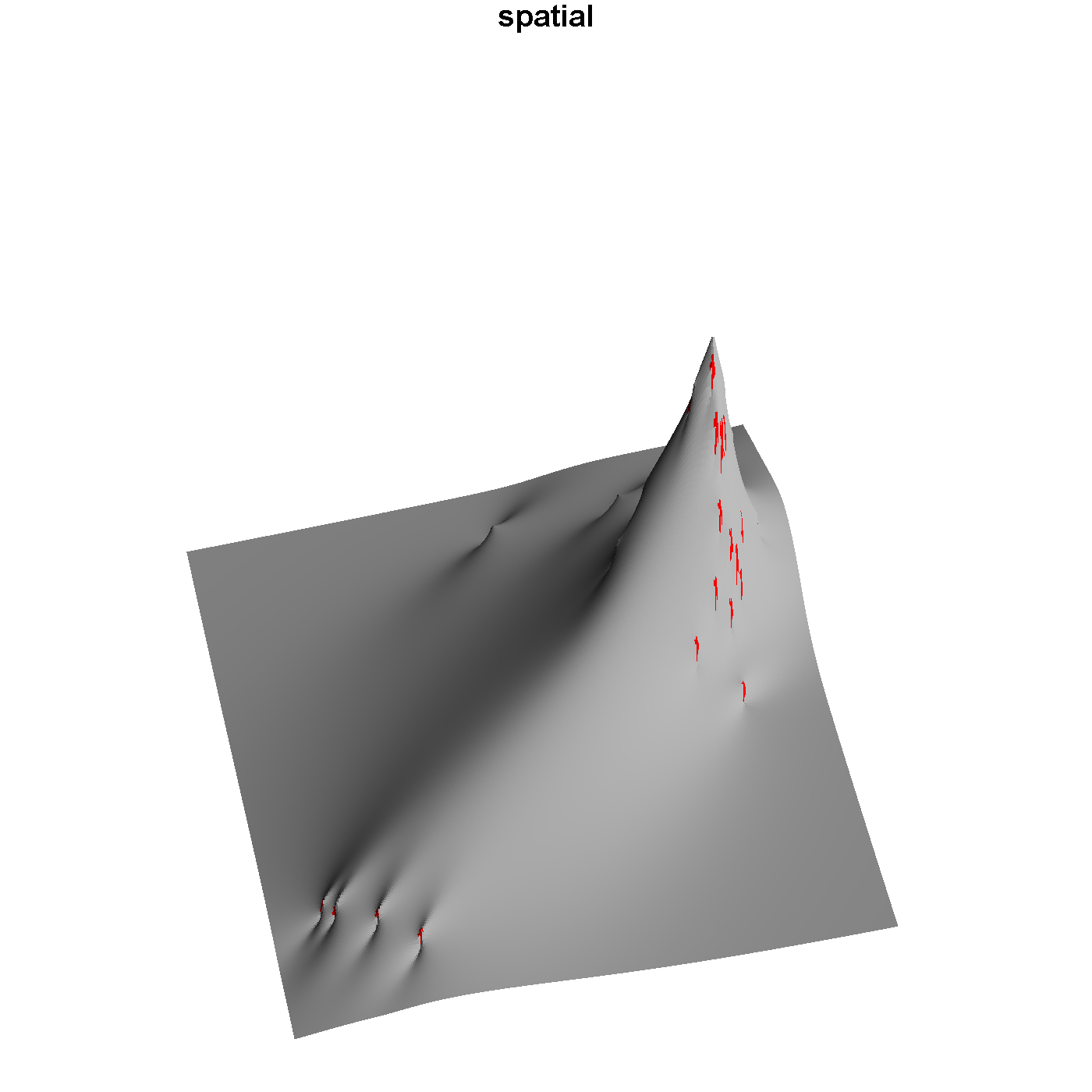}
\hspace{\ws}\includegraphics[keepaspectratio=true,width = \ww, trim = 10mm 16mm 10mm 20mm, clip, page = 5]{depth_contours_noborder.pdf}\\
{\small Spatial depth}}}
         \caption{Depth plots and contours of bivariate data.} 
         \label{fig:bivariate_depths1}
     \end{center}
\end{figure}
\begin{figure}
     \begin{center}
\fbox{\vbox{\hsize=\wb\includegraphics[keepaspectratio=true,width = \ww, trim = 27mm 6.5mm 28mm 54mm, clip,scale=0.80]{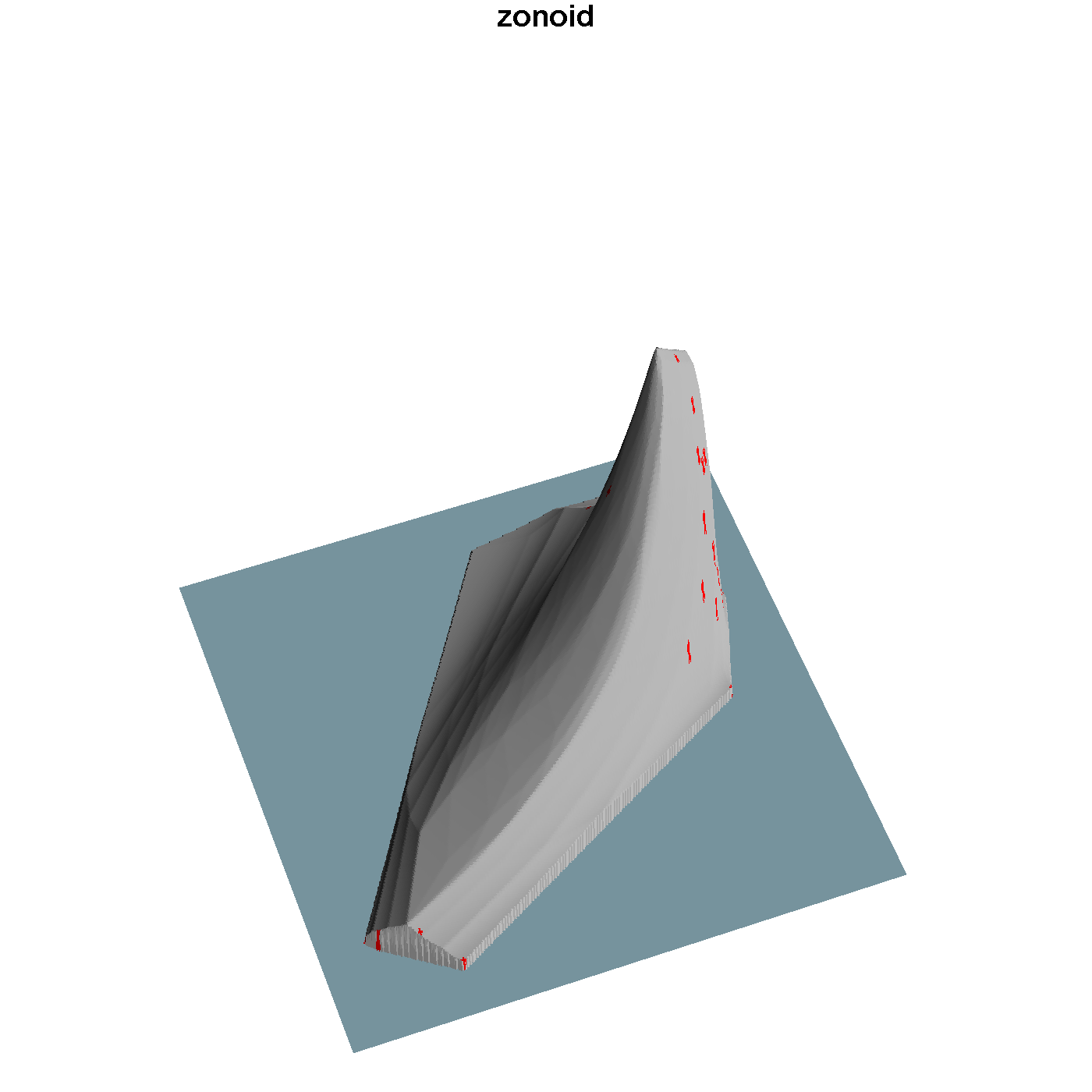}
\hspace{\ws}\includegraphics[keepaspectratio=true,width = \ww, trim = 10mm 16mm 10mm 20mm, clip, page = 2]{depth_contours_noborder.pdf}\\
{\small Zonoid depth}}}
\fbox{\vbox{\hsize=\wb\includegraphics[keepaspectratio=true,width = \ww, trim = 27mm 6.5mm 28mm 54mm, clip,scale=0.80]{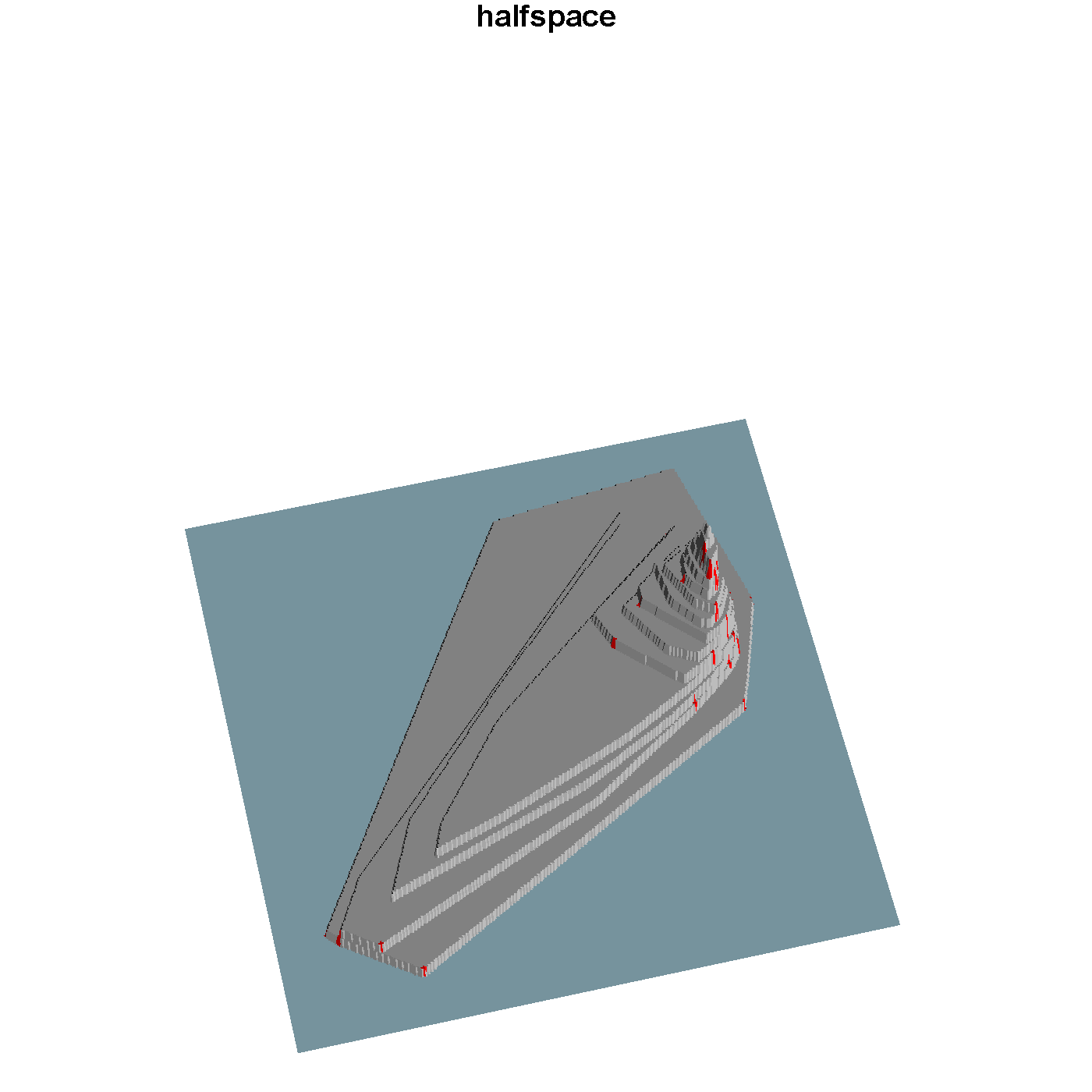}
\hspace{\ws}\includegraphics[keepaspectratio=true,width = \ww, trim = 10mm 16mm 10mm 20mm, clip, page = 8]{depth_contours_noborder.pdf}\\
{\small Halfspace depth}}}
\fbox{\vbox{\hsize=\wb\includegraphics[keepaspectratio=true,width = \ww, trim = 27mm 6.5mm 28mm 54mm, clip,scale=0.80]{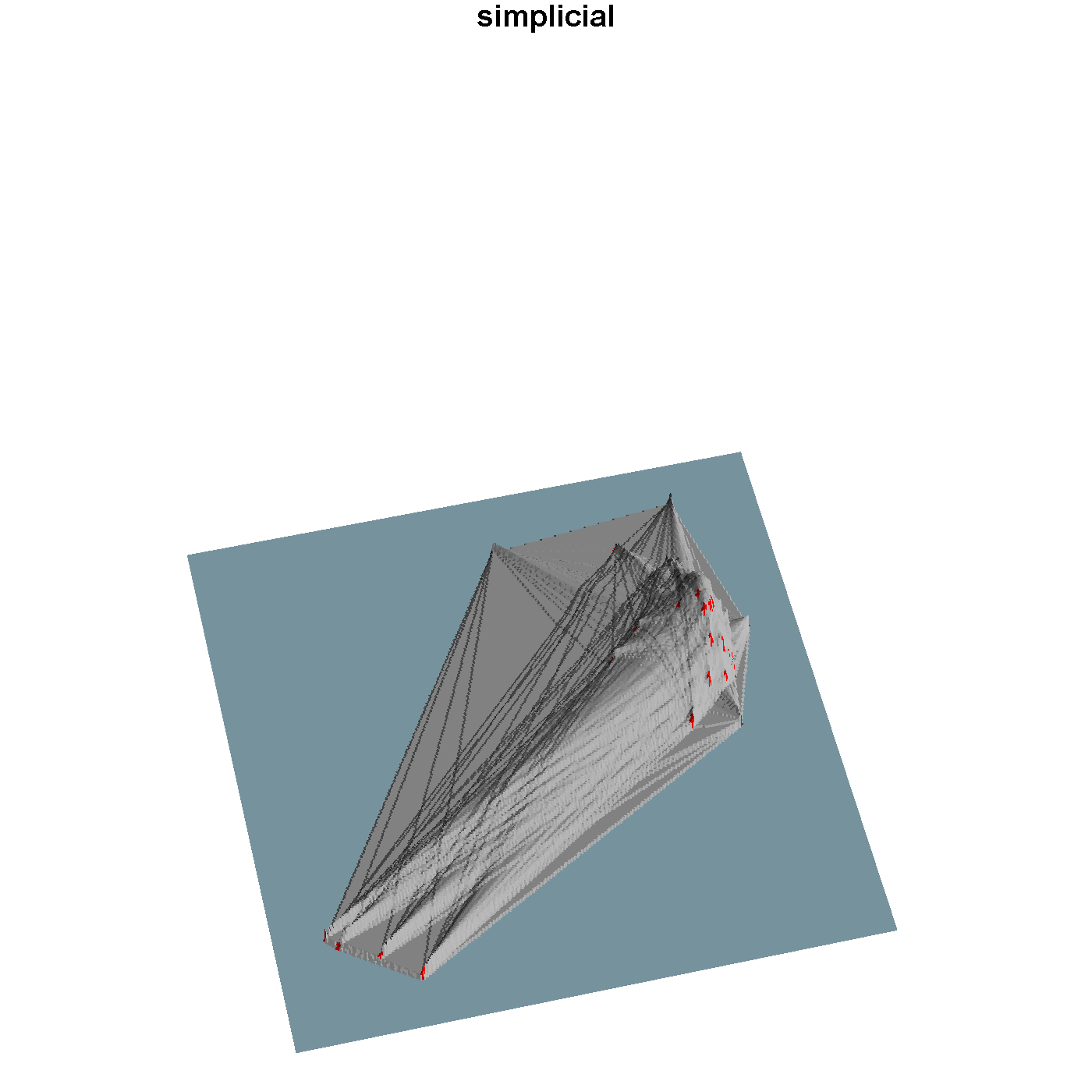}
\hspace{\ws}\includegraphics[keepaspectratio=true,width = \ww, trim = 10mm 16mm 10mm 20mm, clip, page = 6]{depth_contours_noborder.pdf}\\
{\small Simplicial depth}}}
\fbox{\vbox{\hsize=\wb\includegraphics[keepaspectratio=true,width = \ww, trim = 27mm 6.5mm 28mm 54mm, clip,scale=0.80]{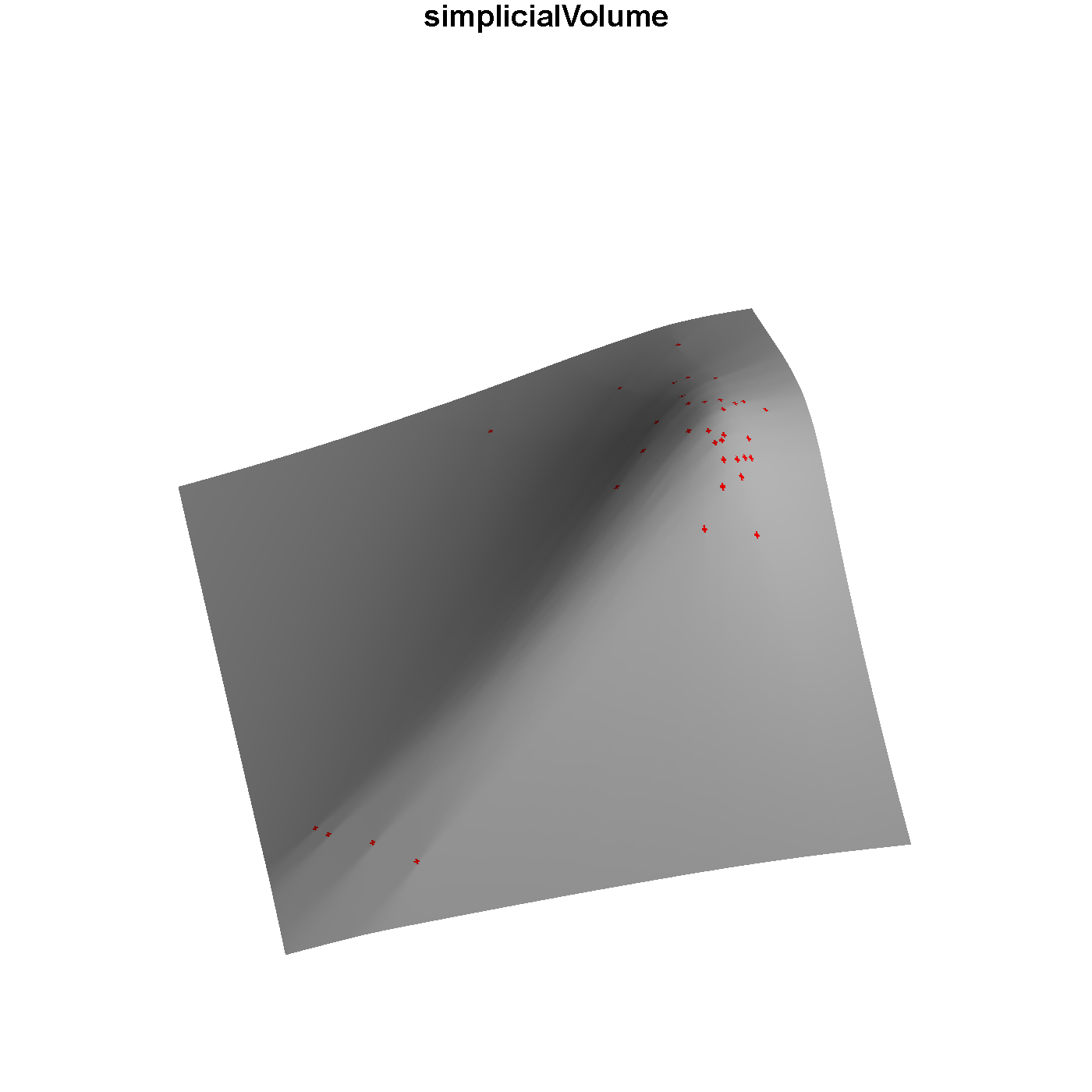}
\hspace{\ws}\includegraphics[keepaspectratio=true,width = \ww, trim = 10mm 16mm 10mm 20mm, clip, page = 7]{depth_contours_noborder.pdf}\\
{\small Simplicial volume depth}}}
         \caption{Depth plots and contours of bivariate data.} 
         \label{fig:bivariate_depths2}
     \end{center}
\end{figure}

\textbf{\emph{Mahalanobis depth}} is based on an outlyingness measure, \textit{viz.} the Mahalanobis distance \citep{Mahalanobis36} between $\boldsymbol{z}$ and a center of $\boldsymbol{X}$, $\boldsymbol{\mu}(\boldsymbol{X})$ say:
\begin{equation*}
    d^2_{Mah}\bigl(\boldsymbol{z};\boldsymbol{\mu}(\boldsymbol{X}),\boldsymbol{\Sigma}(\boldsymbol{X})\bigr)=\bigl(\boldsymbol{z}-\boldsymbol{\mu}(\boldsymbol{X})\bigr)^\prime\boldsymbol{\Sigma}(\boldsymbol{X})^{-1}\bigl(\boldsymbol{z}-\boldsymbol{\mu}(\boldsymbol{X})\bigr).
\end{equation*}
The depth of a point $\boldsymbol{z}$ w.r.t.~$\boldsymbol{X}$ is then defined as \citep{Liu92}
\begin{equation}\label{equ:MahDepth}
    D_{Mah}(\boldsymbol{z}|\boldsymbol{X})=\frac{1}{1+d^2_{Mah}\bigl(\boldsymbol{z};\boldsymbol{\mu}(\boldsymbol{X}),\boldsymbol{\Sigma}(\boldsymbol{X})\bigr)},
\end{equation}
where $\boldsymbol{\mu}(\boldsymbol{X})$ and $\boldsymbol{\Sigma}(\boldsymbol{X})$ are appropriate estimates of mean and covariance of $\boldsymbol{X}$.
This depth function obviously satisfies all the above postulates and is quasi-concave, too. It can be regarded as a \emph{parametric depth} as it is defined by a finite number of parameters (namely $\frac{d(d+1)}{2}$). Based on the two first moments, its depth contours are always ellipsoids centered at $\boldsymbol{\mu}(\boldsymbol{X})$, and thus independent of the shape of $\boldsymbol{X}$. If $\boldsymbol{\mu}(\boldsymbol{X})$ and $\boldsymbol{\Sigma}(\boldsymbol{X})$ are chosen to be moment estimates, \textit{i.e.}~ $\boldsymbol{\mu}(\boldsymbol{X})=\frac{1}{n}\boldsymbol{X}^\prime\boldsymbol{1}_n$ being the traditional \emph{average} and $\boldsymbol{\Sigma}(\boldsymbol{X})=\frac{1}{n-1} (\boldsymbol{X}-\boldsymbol{1}_n\boldsymbol{\mu}(\boldsymbol{X})^\prime)^\prime(\boldsymbol{X}-\boldsymbol{1}_n\boldsymbol{\mu}(\boldsymbol{X})^\prime)$ being the \emph{empirical covariance matrix}, the corresponding depth may be sensitive to outliers. A more robust depth is obtained with the \emph{minimum covariance determinant} (MCD) estimator, see \citet{RousseeuwL87}. 

Calculation of the Mahalanobis depth consists in estimation of the center vector $\boldsymbol{\mu}(\boldsymbol{X})$ and the inverse of the scatter matrix $\boldsymbol{\Sigma}(\boldsymbol{X})$. In the simplest case of traditional moment estimates the time complexity amounts to $O(nd^2 + d^3)$ only. \citet{RousseeuwD99} develop an efficient algorithm for computing robust MCD estimates.

\textbf{\emph{Projection depth}}, similar to Mahalanobis depth, is based on a measure of outlyingness. See \citet{Stahel81}, \citet{Donoho82}, and also \citet{Liu92}, \citet{ZuoS00}. The worst case outlyingness is obtained by maximizing an outlyingness measure over all univariate projections:
\begin{equation*}
    o_{prj}(\boldsymbol{z}|\boldsymbol{X})=\sup_{{\boldsymbol u}\in S^{d-1}}\frac{|\boldsymbol{z}^\prime\boldsymbol{u}-m(\boldsymbol{X}^\prime\boldsymbol{u})|}{\sigma(\boldsymbol{X}^\prime\boldsymbol{u})},
\end{equation*}
with $m(\boldsymbol{y})$ and $\sigma(\boldsymbol{y})$ being any location and scatter estimates of a univariate sample $\boldsymbol{y}$. Taking 
$m(\boldsymbol{y})$ as the mean and $\sigma(\boldsymbol{y})$ as the standard deviation
one gets the Mahalanobis outlyingness, due to the projection property \citep{Dyckerhoff04}. In the literature and in practice most often \emph{median}, $med(\boldsymbol{y})=\boldsymbol{y}_{\left(\frac{\lfloor n+1 \rfloor}{2}\right)}$, and \emph{median absolute deviation from the median}, $MAD(\boldsymbol{y})=med(|\boldsymbol{y}-med(\boldsymbol{y})\boldsymbol{1}_n|)$, are used, as they are robust.
Projection depth is then obtained as
\begin{equation}
    D_{prj}({\boldsymbol z}|\boldsymbol{X})=\frac{1}{1+o_{prj}(\boldsymbol{z}|\boldsymbol{X})}.
\end{equation}

This depth satisfies all the above postulates and quasiconcavity. By involving the symmetric scale factor $MAD$ its contours are centrally symmetric and thus are not well suited for describing skewed data.

Exact computation of the projection depth is a nontrivial task, which fast becomes intractable for large $n$ and $d$. \citet{LiuZ14} suggest an algorithm \citep[and a MATLAB implementation, see][]{LiuZ15}. In practice one may approximate the projection depth from above by minimizing it over projections on $k$ random lines, which has time complexity $O(knd)$.
It can be shown that finding the exact value is a zero-probability event though.

\textbf{\emph{Spatial depth}} (also $L_1$-depth) is a distance-based depth formulated by \citet{VardiZ00} and \citet{Serfling02}, exploiting the idea of spatial quantiles of \citet{Chaudhuri96} and \citet{Koltchinskii97}. 
 For a point $\boldsymbol{z}\in\mathbb{R}^d$, it is defined as one minus the length of the average direction from $\boldsymbol{X}$ to $\boldsymbol{z}$:
\begin{equation}\label{equ:sptDepth}
    D_{spt}(\boldsymbol{z}|\boldsymbol{X})=1-\Bigl\|\frac{1}{n}\sum_{i=1}^n \boldsymbol{v}\bigl(\boldsymbol{\Sigma}^{-\frac{1}{2}}(\boldsymbol{X})(\boldsymbol{z} - \boldsymbol{x}_i)\bigr)\Bigr\|,
\end{equation}
with $\boldsymbol{v}(\boldsymbol{y})=\frac{\boldsymbol{y}}{\|\boldsymbol{y}\|}$ if $\boldsymbol{y}\neq\boldsymbol{0}$, and $\boldsymbol{v}(\boldsymbol{0})=\boldsymbol{0}$.
The scatter matrix $\boldsymbol{\Sigma}(\boldsymbol{X})$ provides the affine invariance.

Affine invariant spatial depth satisfies all the above postulates, but is not quasiconcave. Its maximum is referred to as the \emph{spatial median}. In the one-dimensional case it coincides with the halfspace depth, defined below.

Spatial depth can be efficiently computed even for large samples amounting in the simplest case to time complexity $O(nd^2 + d^3)$; for calculation of $\boldsymbol{\Sigma}^{-\frac{1}{2}}(\boldsymbol{X})$ see the above discussion of the Mahalanobis depth.

\textbf{\emph{Halfspace depth}} follows the idea of \citet{Tukey75}, see also \citet{DonohoG92}. The Tukey (=halfspace, location) depth of $\boldsymbol{z}$ w.r.t.~$\boldsymbol{X}$ is determined as:
\begin{equation}
    D_{hs}(\boldsymbol{z}|\boldsymbol{X}) = \min_{\boldsymbol{u}\in S^{d-1}}\frac{1}{n}\#\{i:\boldsymbol{x}_i^\prime\boldsymbol{u}\le\boldsymbol{z}^\prime\boldsymbol{u};i=1,...,n\}.
\end{equation}

Halfspace depth satisfies all the postulates of a depth function. In addition, it is quasiconcave, and equals zero outside the convex hull of the support of $\boldsymbol{X}$. For any $\boldsymbol{X}$, there exists at least one point having depth not smaller than $\frac{1}{1+d}$ \citep{Mizera02}. For empirical distributions, halfspace depth is a discrete function of ${\boldsymbol z}$, and the set of depth-maximizing locations --- the \emph{halfspace median} --- can consist of more than one point (to obtain a unique median, an average of this deepest trimmed region can be calculated). Halfspace depth determines the empirical distribution uniquely \citep{StruyfR99,Koshevoy02}.

\citet{DyckerhoffM16} develop a family of algorithms (for each $d>1$) possessing time complexity $O(n^{d-1}\log{n})$ and $O(n^d)$ (the last has proven to be computationally more efficient for larger $d$ and small $n$). These algorithms are applicable for moderate $n$ and $d$. For large $n$ or $d$ and (or) if the depth has to be computed many times, approximation by minimizing over projections on random lines can be performed \citep{Dyckerhoff04,CuestaAlbertosNR08}. By that, $D_{hs}(\boldsymbol{z}|\boldsymbol{X})$ is approximated from above with time complexity $O(knd)$, and $D_{hs}(\boldsymbol{X}|\boldsymbol{X})$ with time complexity $O\bigl(kn(d + \log{n})\bigr)$, using $k$ random directions \citep[see also][]{MozharovskyiML15}.

\textbf{\emph{Simplicial depth}} \citep{Liu90}
is defined as the portion of simplices having vertices from $\boldsymbol{X}$ which contain $\boldsymbol{z}$:
\begin{equation}
    D_{sim}(\boldsymbol{z}|\boldsymbol{X}) = \frac{1}{{n \choose d+1}}\sum_{1\le i_1<i_2<...<i_{d+1}\le n}I\bigl(\boldsymbol{z}\in\text{conv}(\boldsymbol{x}_{i_1},\boldsymbol{x}_{i_2},...,\boldsymbol{x}_{i_{d+1}})\bigr)
\end{equation}
with $\text{conv}(\mathcal{Y})$ being the convex hull of $\mathcal{Y}$ and $I(\mathcal{Y})$ standing for the indicator function, which equals $1$ if $\mathcal{Y}$ is true and $0$ otherwise. 

It satisfies postulates (D1), (D2), (D3), and (D5). The set of depth-maximizing locations is not a singleton, but, different to the halfspace depth, it is not convex (in fact it is not even necessarily connected) and thus simplicial depths fails to satisfy (D4). It characterizes the empirical measure if the data, \textit{i.e.}~the rows of $\boldsymbol{X}$, are in general position, and is, as well as the halfspace depth, due to its nature rather insensitive to outliers, but vanishes beyond the convex hull of the data $\text{conv}(\boldsymbol{X})$.

Exact computation of the simplicial depth has time complexity of $O(n^{d+1}d^3)$. Approximations accounting for a part of simplices can lead to time complexity $O(kd^3)$ only when drawing $k$ random $(d+1)$-tuples from $\boldsymbol{X}$, or reduce real computational burden with the same time complexity, but keeping precision when drawing a constant portion of ${n \choose d+1}$.
For $\mathbb{R}^2$, \cite{RousseeuwR96} proposed an exact efficient algorithm with time complexity $O(n\log n)$ .

\textbf{\emph{Simplicial volume depth}} \citep{Oja83} is defined via the average volume of the simplex with $d$ vertices from $\boldsymbol{X}$ and one being $\boldsymbol{z}$:
\begin{equation}
    D_{simv}(\boldsymbol{z}|\boldsymbol{X}) = \frac{1}{1 + \frac{1}{{n \choose d}\sqrt{\text{det}\bigl(\boldsymbol{\Sigma}(\boldsymbol{X})\bigr)}}\sum_{1\le i_1<i_2<...<i_{d}\le n}\text{vol}\bigl(\text{conv}(\boldsymbol{z},\boldsymbol{x}_{i_1},\boldsymbol{x}_{i_2},...,\boldsymbol{x}_{i_{d}})\bigr)}
\end{equation}
with $\text{vol}(\mathcal{Y})$ being the Lebesgue measure of $\mathcal{Y}$.

It satisfies all above postulates, is quasiconcave, determines $\boldsymbol{X}$ uniquely \citep{Koshevoy03}, and has a nonunique median.

Time complexity of the exact computation of the simplicial volume depth amounts to $O(n^d d^3)$, and thus approximations similar to the simplicial depth may be necessary.

\textbf{\emph{Zonoid depth}} has been first introduced by \citet{KoshevoyM97}, see also \citet{Mosler02} for a discussion in detail. 
The zonoid depth function is most simply defined by means of depth contours --- the zonoid trimmed regions. The zonoid $\alpha$-trimmed region of an empirical distribution is defined as follows: For $\alpha\in\left[\frac{k}{n},\frac{k+1}{n}\right],\,k=1,...,n-1$ the zonoid region is defined as
\begin{equation*}
    Z_{\alpha}(\boldsymbol{X})=\mbox{conv}\Bigl\{\frac{1}{\alpha n}\sum_{j=1}^k{\boldsymbol x}_{i_j}+\Bigl(1-\frac{k}{\alpha n}\Bigr){\boldsymbol x}_{i_{k+1}}:\{i_1,...,i_{k+1}\}\subset \{1,...,n\}\Bigr\},
\end{equation*}
and for $\alpha\in\left[0,\frac{1}{n}\right)$
\begin{equation*}
    Z_{\alpha}(\boldsymbol{X})=\mbox{conv}(\boldsymbol{X}).
\end{equation*}
Thus, \textit{e.g.}, $Z_{\frac{3}{n}}(\boldsymbol{X})$ is the convex hull of the set of all possible averages involving three points of $\boldsymbol{X}$, and $Z_0(\boldsymbol{X})$ is just the convex hull of $\boldsymbol{X}$.

The zonoid depth of a point ${\boldsymbol z}$ w.r.t.~$\boldsymbol{X}$ is then defined as the largest $\alpha\in[0,1]$ such that $Z_{\alpha}(\boldsymbol{X})$ contains ${\boldsymbol z}$ if ${\boldsymbol z}\in\mbox{conv}(\boldsymbol{X})$ and $0$ otherwise:
\begin{equation}
    D_{zon}(\boldsymbol{z}|\boldsymbol{X})=
        \sup\{\alpha\in[0,1]:\,\boldsymbol{z}\in Z_{\alpha}(\boldsymbol{X})\},
\end{equation}
where $\sup$ of $\emptyset$ is defined to be 0.

The zonoid depth belongs to the class of weighted-mean depths, see \citet{DyckerhoffM11}. It satisfies all the above postulates and is quasiconcave. As well as halfspace and simplicial depth, zonoid depth vanishes beyond the convex hull of $\boldsymbol{X}$. Its maximum (always equaling $1$) is located at the mean of the data, thus this depth is not robust.


Its exact computation with the algorithm of \citet{DyckerhoffKM96}, based on linear programming and exploiting the idea of Danzig-Wolf decomposition, appears to be fast enough for large $n$ and $d$, not to need approximation.

A common property of the considered above depth notions is that they concentrate on global features of the data ignoring local specifics of sample geometry.
Thus they are unable to reflect multimodality of the underlying distribution. Several depths have been proposed in the literature to overcome this difficulty. Two of them were introduced in the classification context, localized extension of the spatial depth \citep{DuttaG15} and the data potential \citep{PokotyloM16}. They are also implemented in the \proglang{R}-package \pkg{ddalpha}. The performance of these depths and of the classifiers exploiting them depends on the type of the kernel and its bandwidth. While the behaviour of these two notions substantially differs from the seven depth notions mentioned above, we leave them beyond the scope of this article and relegate to the corresponding literature for theoretical and experimental results.


\subsection{Computation time}\label{ssec:depthComputation}

\begin{figure}[!h]
  \begin{center}
    \includegraphics[keepaspectratio=true,width = \textwnew, trim = 3mm 2mm 0mm 2mm, clip, page = 1]{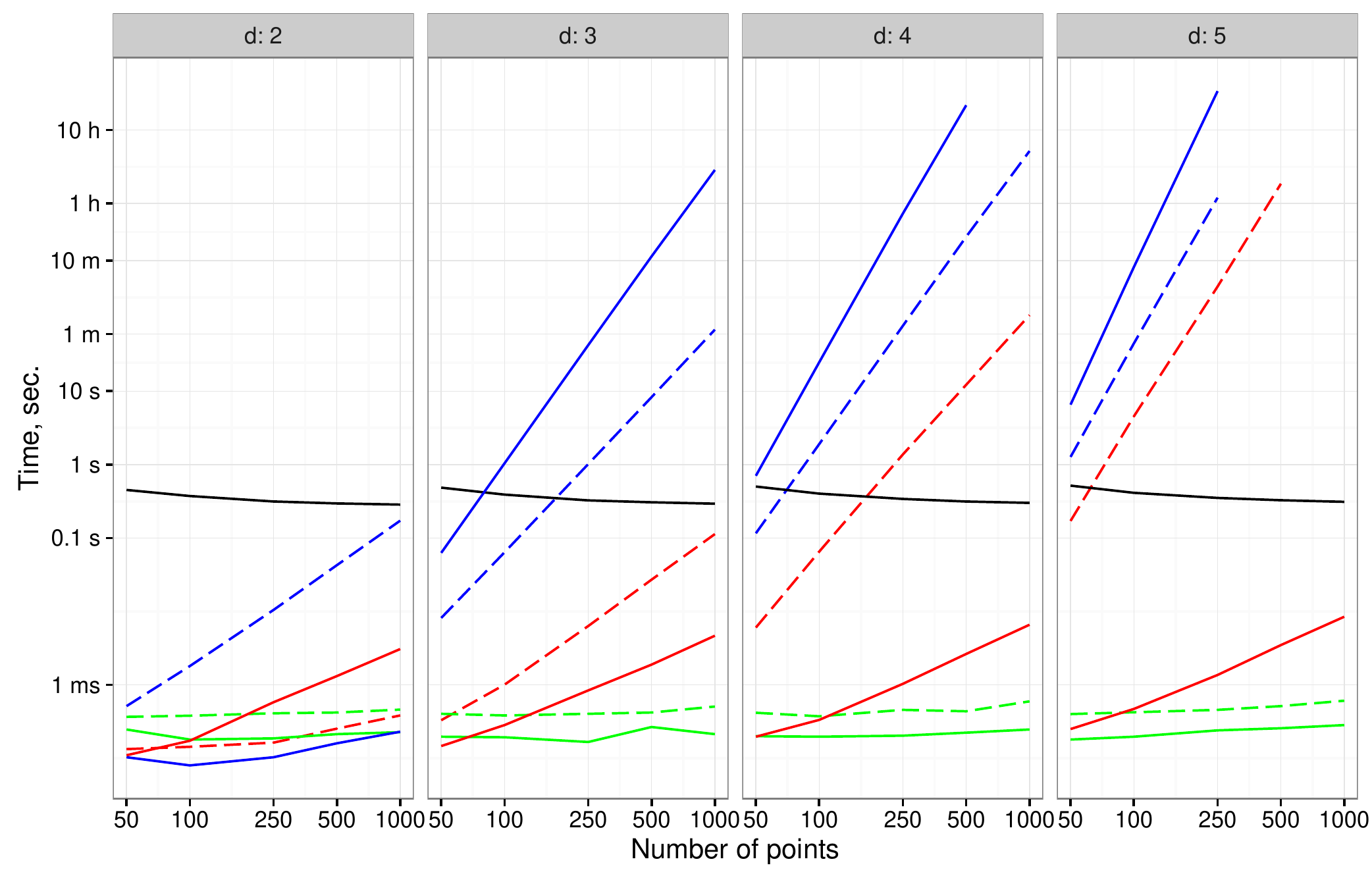}\\
    {\footnotesize{\color{red}\LBL} zonoid, {\color{red}\LBD} halfspace, {\color{green}\LBL} Mahalanobis, {\color{green}\LBD} spatial, {\color{black}\LBL} projection, {\color{blue}\LBL} simplicial, {\color{blue}\LBD} simplicial volume}
    \caption{Calculation time of various depth functions, on the logarithmic time scale.}
    \label{fig:depth_speed}
  \end{center}
\end{figure}

To give insights into the speed of exactly calculating various depth notions we indicate computation times by graphics in Figure~\ref{fig:depth_speed}. On the logarithmic time scale, the lines represent the time (in seconds) needed to compute the depth of a single point, averaged over 50 points w.r.t.~60 samples, varying dimension $d\in\{2, 3, 4, 5\}$ and sample length $n\in\{50,100,250,500,1000\}$.
Due to the fact that computation times of the algorithms do not depend on the particular shape of the data, the data has been drawn from the standard normal distribution. Some of the graphics are incomplete due to excessive time. Projection depth has been approximated using 10\,000\,000$/ n$ random projections, all other depths have been computed exactly. Here we used one kernel of the Intel Core i7-4770 (3.4 GHz) processor having enough physical
memory.

One can see that, for all considered depths and $n\le 1\,000$, computation of the two-dimensional depth never oversteps one second.
For halfspace and simplicial depth this can be explained by the fact that in the bivariate case both depths depend only on the angles between the lines connecting $\boldsymbol z$ with the data points $\boldsymbol x_i$ and the abscissa. Computing these angles and sorting them has a complexity of $O(n\log n)$ which determines the complexity of the bivariate algorithms.
As expected, halfspace, simplicial, and simplicial volume depths, being of combinatorial nature, have exponential time growth in $(n,d)$. Somewhat surprising, zonoid depth being computed by linear programming, seems to be way less sensitive to dimension. One can conclude that in applications with restricted computational resources, halfspace, projection, simplicial and simplicial volume depths may be rather approximated in higher dimensions, while exact algorithms can still be used in the low-dimensional framework, \textit{e.g.} when computing time cuts of multivariate functional depths, or to assess the performance of approximation algorithms.

\subsection{Maximum depth classifier}\label{ssec:depthMaxdepth}
\begin{figure}[!h]
  \begin{center}
    \includegraphics[keepaspectratio=true,width = \textwnew, trim = 3mm 2mm 0mm 2mm, clip, page = 1]{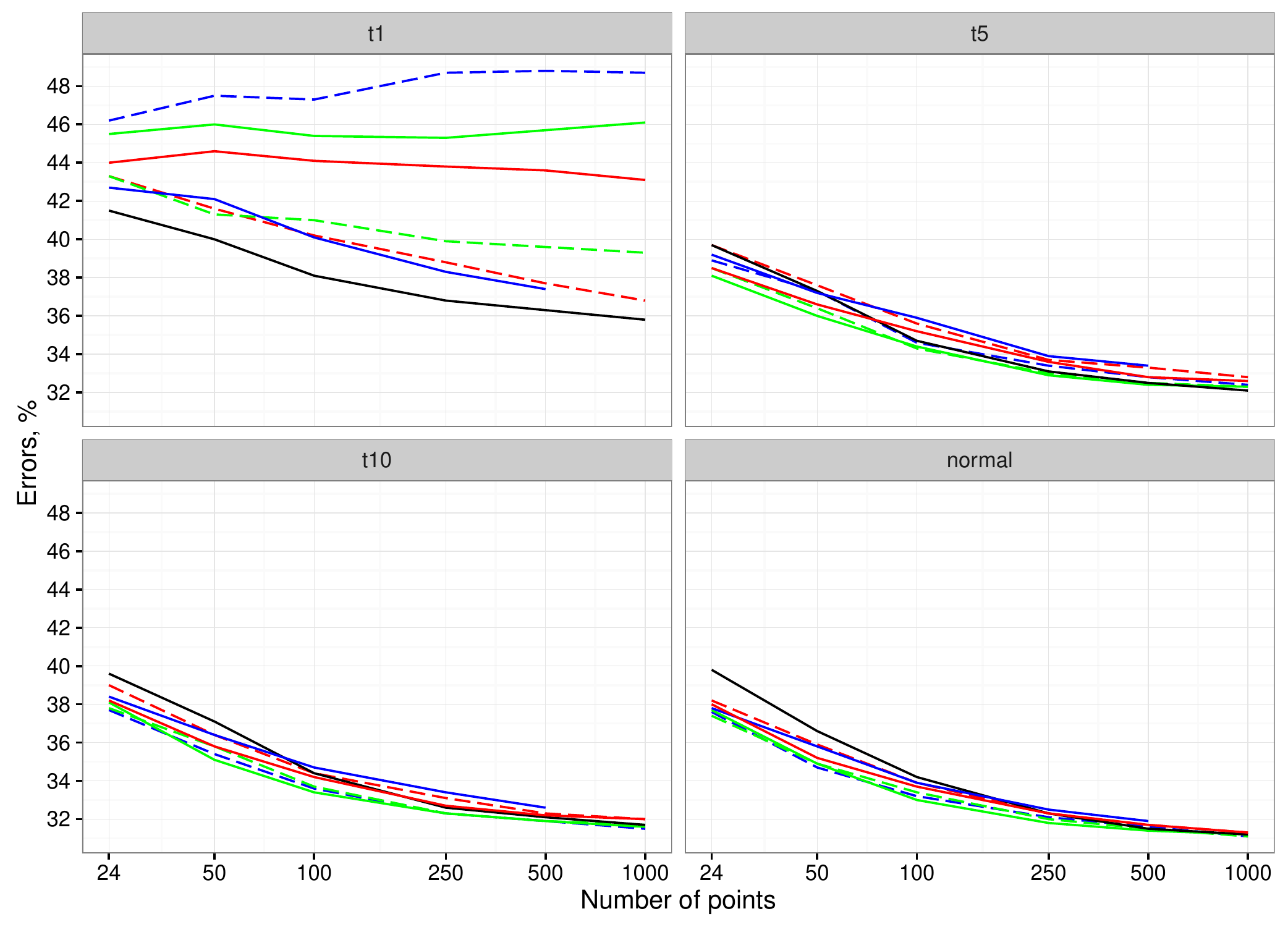}\\
    {\footnotesize{\color{red}\LBL} zonoid, {\color{red}\LBD} halfspace, {\color{green}\LBL} Mahalanobis, {\color{green}\LBD} spatial, {\color{black}\LBL} projection, {\color{blue}\LBL} simplicial, {\color{blue}\LBD} simplicial volume}
    \caption{Average error rates of the maximum depth classifier with different data depths. The samples are simulated from the Student-$t$ distribution possessing $1$, $5$, $10$, and $\infty$ degrees of freedom.}
    \label{fig:maxdepthgraphs}
  \end{center}
\end{figure}

To demonstrate the differing finite-sample behavior of the above depth notions and to construct a bridge to supervised classification, in this section we compare the depths in the frame of the maximum depth classifier. This is obtained by simply choosing the class in which $\boldsymbol{x}_0$ has the highest depth (breaking ties at random):
\begin{equation}
	class(\boldsymbol{x}_0) = \argmax_{i\in\{1,...,q\}}\,D(\boldsymbol{x}_0|\boldsymbol{X_i}).
\end{equation}
\cite{GhoshC05b} have proven that its misclassification rate converges to the optimal Bayes risk if each $\boldsymbol{X}_i,\,i=1,...,q$, is sampled from a unimodal elliptically symmetric distribution having a common nonincreasing density function, a prior probability $\frac{1}{q}$, and differing in location parameter only (location-shift model), for halfspace, simplicial, and projection depths, and under additional assumptions for spatial and simplicial volume depths. Setting $q=2$, and $n=24,50,100,250,500,1000$, $n_i=n/2,\,i=1,2$, we sample $\boldsymbol{X}_i$ from a Student-$t$ distribution with location parameters $\mu_1=[0,0]$, $\mu_2=[1,1]$ and common scale parameter
$\Sigma=\left[ \begin{smallmatrix}  1 & 1 \\  1 & 4 \\ \end{smallmatrix}\right]$, setting the degrees of freedom to $t=1, 5, 10, \infty$. Average error rates over 250 samples each checked on $1000$ observations are indicated in Figure~\ref{fig:maxdepthgraphs}. The testing observations were sampled inside the convex hull of the training set. The problem of outsiders is addressed in Section~\ref{sec:outsiders}. For $n=1000$, experiments have not been conducted with the simplicial depth due to high computation time.

As expected, with increasing $n$ and $t$ classification error and difference between various depths decrease. As the classes stem from elliptical family, depths accounting explicitly for ellipticity (Mahalanobis and spatial due to covariance matrix), symmetry of the data (projection), and also volume, form the error frontier. On the other hand, except for the projection depth, they are nonrobust and perform poorly for Cauchy distribution. While projection depth, even being approximated, behaves excellent in all the experiments, it may perform poorly if distributions of $\boldsymbol{X}_i$ retain asymmetry due to inability to reflect this.

\section[Classification in the $DD$-plot]{Classification in the $DD$-plot}\label{sec:ddclass}

In Section~\ref{ssec:depthMaxdepth}, we have already considered the naive way of depth-based classification --- the maximum depth classifier. Its extension beyond the equal-prior location-shift model, \textit{e.g.} to account for differing shape matrices of the two classes, or unequal prior probabilities, is somewhat cumbersome, \textit{cf.}~\cite{GhoshC05a,CuiLY08}. A simpler way, namely to use the $DD$-plot (or, more general, a $q$-dimensional depth space), has been proposed by~\cite{LiCAL12}. For a training sample consisting of $\boldsymbol{X}_1,...,\boldsymbol{X}_q$, the depth space is constructed by applying the mapping $\mathbb{R}^d\rightarrow[0,1]^q\,:\,\boldsymbol{x}\mapsto\bigl(D(\boldsymbol{x}|\boldsymbol{X}_1),...,D(\boldsymbol{x}|\boldsymbol{X}_q)\bigr)$ to each of the observations. Then the classification is performed in this low-dimensional space of depth-extracted information, which, \textit{e.g.}, for $q=2$ is just a unit square. The core idea of the $DD\alpha$-classifier is the $DD\alpha$-separator, a fast heuristic for the $DD$-plot. This is presented in Section~\ref{ssec:ddclassAlpha}, where we slightly abuse the notation introduced before. This is done in an intuitive way for the sake of understandability and closeness to the implementation. Further, in Section~\ref{ssec:ddclassAlternatives} we discuss application of alternative techniques in the depth space.

\subsection[The DDalpha-separator]{The $DD\alpha$-separator}\label{ssec:ddclassAlpha}

The $DD\alpha$-separator is an extension of the $\alpha$-procedure to the depth space, see \cite{Vasilev03,VasilevL98}, also \cite{LangeM14}. It iteratively synthesizes the space of features, coordinate axes of the depth space or their (polynomial) extensions, choosing features minimizing a two-dimensional empirical risk in each step. The process of space enlargement stops when adding features does not further reduce the empirical risk. Here we give its comprehensive description. The detailed algorithm is stated right below.

\begin{figure}[!h]
  \renewcommand{\trimval}{0 17 29 58}
  \begin{center}
    \includegraphics[keepaspectratio=true,width = .325\textwnew, page = 1, Trim=\trimval, clip]{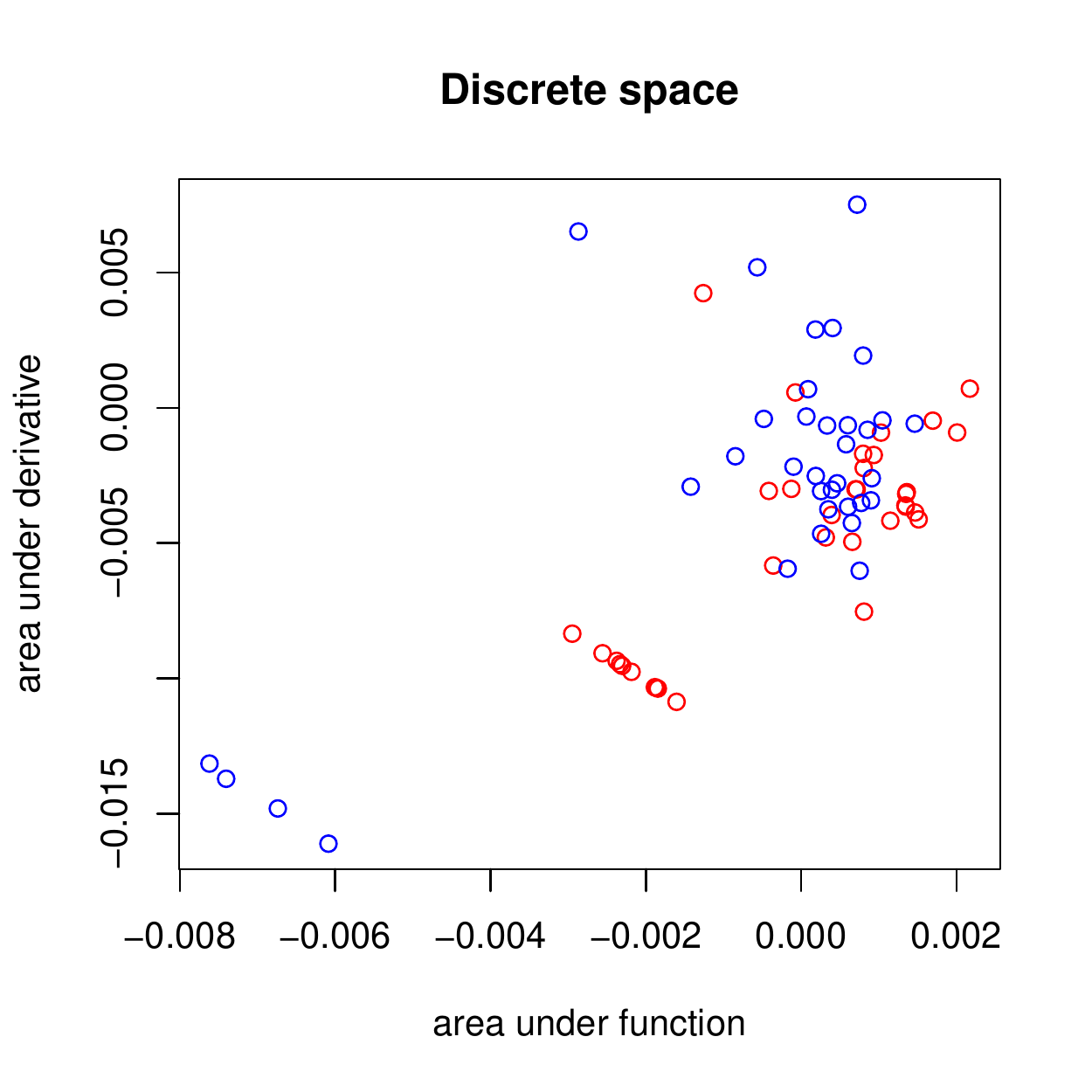}
    \includegraphics[keepaspectratio=true,width = .325\textwnew, page = 4, Trim=\trimval, clip]{ddaplha_steps_discrete_space.pdf}
    \includegraphics[keepaspectratio=true,width = .325\textwnew, page = 46, Trim=\trimval, clip]{ddaplha_steps.pdf}
    \caption{The discretized space (left), the depth contours with the separating rule (middle) and the $DD$-plot with the separating line in it (right), using spatial depth. Here
    we denote the depth of a point w.r.t.~red and blue classes by $x$ and $y$, respectively.}
    \label{fig:discretespace}
    \end{center}
\end{figure}

Regard the two-class sample illustrated on Figure~\ref{fig:discretespace}, left, representing discretizations of the electrocardiogram curves. Explanation of the data is given in Section~\ref{ssec:introOutline}, we postpone the explanation of the discretization scheme till Section~\ref{sec:functions} and consider a binary classification in the $DD$-plot for the moment. Figure~\ref{fig:discretespace}, middle, represents the depth contours of each class computed using the spatial depth. The $DD$-plot is obtained as a depth mapping $(\boldsymbol{X}_1,\boldsymbol{X}_2)\mapsto\boldsymbol{Z}=\{\boldsymbol{z}_i=(D_{i,1},D_{i,2}),\,i=1,...,n_1+n_2\}$,
 when the first class is indexed by $i=1,...,n_1$ and the second by $i=n_1+1,...,n_2$,
 and writing $D_{spt}(\boldsymbol{x}_i|\boldsymbol{X}_1)$ (respectively $D_{spt}(\boldsymbol{x}_i|\boldsymbol{X}_2)$) by $D_{i,1}$ (respectively $D_{i,2}$) for shortness. Further, to enable for nonlinear separation in the depth space, but to employ linear discrimination in the synthesized subspaces, the kernel trick is applied. As the $DD\alpha$-separator explicitly works with the dimensions (space axis), a finite-dimensional resulting space is required.
We choose the space extension degree by means of a fast cross-validation,
which is performed over a small range and in the depth space only. The high computation speed of the $DD\alpha$-separator allows for this.

We use polynomial extension of degree $p$, which results in $r = {p + q\choose q} - 1$ dimensions (by default, we choose $p$ among $\{1,2,3\}$ using 10-fold cross-validation); truncated series or another finitized basis of general reproducing kernel Hilbert spaces can be used alternatively. This extended depth space serves as the input to the $DD\alpha$-separator. For $q = 2$, and taking $p = 3$, one gets the extended depth space $\boldsymbol{Z}^{(p)}$ consisting of observations
$\boldsymbol{z}^{(p)}_i=(D_{i,1},~~D_{i,2},~~D_{i,1}^2,~~D_{i,1}\times D_{i,2},~~D_{i,2}^2,~~D_{i,1}^3,~~D_{i,1}^2\times D_{i,2},~~D_{i,1}\times D_{i,2}^2,~~D_{i,2}^3)\in\mathbb{R}^r$.

After initializations, on the \emph{1st step}, the $DD\alpha$-separator starts with choosing the pair of extended properties minimizing the empirical risk.
For this, it searches through all coordinate subspaces $\boldsymbol{Z}^{(k,l)}=\{\boldsymbol{z}^{(k,l)}_i\,|\,\boldsymbol{z}^{(k,l)}_i=(\boldsymbol{z}^{(p)}_{ik},\boldsymbol{z}^{(p)}_{il}),\,i=1,...,n_1+n_2\}$ for all $\,1\le k<l \le r$, \textit{i.e.}~all pairs of coordinate axis of $\boldsymbol{Z}^{(p)}$.
For each of them, the angle $\alpha_1^{(k,l)}$ minimizing the empirical risk is found
\begin{equation}\label{equ:alphaAlpha}
    \alpha_1^{(k,l)} \in \argmin_{\alpha\in[0;2\pi)} \Delta^{(k,l)}(\alpha)
\end{equation}
with
\begin{equation}\label{equ:alphaPlane}
    \Delta^{(k,l)}(\alpha) = \sum_{i=1}^{n_1} I(\boldsymbol{z}^{(p)}_{ik}\cos\alpha - \boldsymbol{z}^{(p)}_{il}\sin\alpha < 0) + \sum_{i = n_1 + 1}^{n_1 + n_2} I(\boldsymbol{z}^{(p)}_{ik}\cos\alpha - \boldsymbol{z}^{(p)}_{il}\sin\alpha > 0).
\end{equation}
For the regarded example, this is demonstrated in Figure~\ref{tab:gt} by the upper triangle of the considered subspaces. Computationally, it is reasonable to check only those $\alpha$ corresponding to (radial) intervals between points and to choose $\alpha_1^{(k,l)}$ as an average angle between two points from $\boldsymbol{Z}^{(k,l)}$ in case there is a choice, as it is implemented in procedure \textsl{GetMinError}.
Computational demand is further reduced by skipping uninformative pairs, \textit{e.g.}, if one feature is a power of another one and, therefore, the bivariate plot 
is collapsed to a line, as shown in Figure~\ref{tab:gt}. Finally, a triplet is chosen:
\begin{equation}\label{equ:alphaRisk}
    (\alpha_1^{(k^*,l^*)},k^*,l^*) \in \argmin_{1\le k<l \le r,\,\alpha\in[0;2\pi)} \Delta^{(k,l)}(\alpha),
\end{equation}
\textit{i.e.}~a two-dimensional coordinate subspace $\boldsymbol{Z}^{(k^*,l^*)}$ in which the minimal empirical risk over all such subspaces is achieved, and the corresponding angle $\alpha_1^{(k^*,l^*)}$ minimizing this. Among all the minimizing triplets (there may be several as empirical risk is discrete) it is reasonable to choose $k^*$ and $l^*$ with the smallest polynomial degree, the simplest model. Using $\alpha_1^{(k^*,l^*)}$, $\boldsymbol{Z}^{(k^*,l^*)}$ is convoluted to a real line
\begin{equation}\label{equ:alphaConvolute}
\boldsymbol{z}^{(1^*)}=\{z_i\,|\,z_i=\boldsymbol{z}^{(p)}_{ik^*}\cos\alpha_1^{(k^*,l^*)} - \boldsymbol{z}^{(p)}_{il^*}\sin\alpha_1^{(k^*,l^*)},\,i=1,...,n_1+n_2\},
\end{equation}
--- first feature of the synthesized space.

On each following $s$\emph{-step} ($s\ge 2$), the $DD\alpha$-separator proceeds as follows. The feature, obtained by the convolution on the previous $(s-1)$\emph{-step}, is coupled with each of the extended properties  of the depth space, such that a space $\boldsymbol{Z}^{((s-1)^*,k)}=\{\boldsymbol{z}^{((s-1)^*,k)}_i\,|\,\boldsymbol{z}^{((s-1)^*,k)}_i=(\boldsymbol{z}^{((s-1)^*)}_{i},\boldsymbol{z}^{(p)}_{ik})$, $i=1,...,n_1+n_2\}$ is regarded, for all $k$ used in no convolution before. For each $\boldsymbol{Z}^{((s-1)^*,k)}$, $\Delta^{((s - 1)^*,k)}(\alpha_s^{(k)})$ and the corresponding empirical-risk-minimizing angle $\alpha_s^{(k)}$ are obtained using (\ref{equ:alphaAlpha}) and (\ref{equ:alphaPlane}). Out of all considered $k$, the one minimizing $\Delta^{((s - 1)^*,k)}(\alpha_s^{(k)})$ is chosen, as in (\ref{equ:alphaRisk}), and the corresponding $\boldsymbol{Z}^{((s-1)^*,k)}$ is convoluted to $\boldsymbol{z}^{(s^*)}$, as in (\ref{equ:alphaConvolute}).
The second part of Figure~\ref{tab:gt} illustrates a possible second step of the algorithm.

Here we present the algorithm of the $DD\alpha$-separator:

\bigskip

\textbf{The main procedure}

Input: $\tilde{\boldsymbol{X}} = \{\tilde{\boldsymbol{x}}_1,...,\tilde{\boldsymbol{x}}_n\}$, $\tilde{\boldsymbol{x}}_i\in\mathbb{R}^d$,
 \\ \phantom{tabInput:s} $\{ y_1,..., y_n\}$, $y_i\in\{-1,1\}$ for all $i=1,...,m=m_{-1}+m_{+1}$.
\begin{enumerate}
    \item ${\boldsymbol{X}} = {\tilde{\boldsymbol{X}}}^T = \{{\boldsymbol{x}}_1,...,{\boldsymbol{x}}_d\}$, ${\boldsymbol{x}}_i\in\mathbb{R}^n$.
    \item Initialize arrays:
        \begin{algenum}
         \item array of available properties $\boldsymbol{P} \gets \{1..d\}$;
         \item array of constructed features $\boldsymbol{F} \gets \emptyset$;
         \item for a feature $f \in \boldsymbol{F}$ denote $f.p$ and $f.\alpha$ the number of the used property and the optimal angle.
        \end{algenum}
    \item \emph{1st step:} Find the first features:
    \begin{algenum}
        \item select optimal starting features considering all pairs from $\boldsymbol{P}$:
        \item[] $(opt_1, opt_2, e_{min}, \alpha) = \arg\min_{g \in \boldsymbol{G}}g.e$ with \item[]
        \mbox{ $\boldsymbol{G} = \{(p_1, p_2, e, \alpha):
                        (e, \alpha)=\text{\textsl{GetMinError}}({\boldsymbol{x}}_{p_1}, {\boldsymbol{x}}_{p_2}),
                        p_1, p_2 \in \boldsymbol{P}, p_1 < p_2 \}$}

        \item $\boldsymbol{F} \gets \boldsymbol{F} \cup \{(opt_1,0), (opt_2,\alpha)\}$
        \item $\boldsymbol{P} \gets \boldsymbol{P} \setminus \{opt_1, opt_2\}$
        \item set current feature $f^\prime = {\boldsymbol{x}}_{opt_1}\times\cos(\alpha) + {\boldsymbol{x}}_{opt_2}\times\sin(\alpha)$
    \end{algenum}

    \item \emph{Following steps:} Search an optimal feature space while empirical error rate decreases\\
    {\bf while} $e_{min}\ne 0$ and $\boldsymbol{P}\ne\emptyset$ {\bf do}
        \begin{algenum}
            \item select next optimal feature considering all properties from $\boldsymbol{P}$:
            \item[] $(opt, \tilde{e}_{min}, \alpha) = \arg\min_{g \in \boldsymbol{G}}g.e$ with \item[]
            $\boldsymbol{G} = \{(p, e, \alpha):
                        (e, \alpha)=\text{\textsl{GetMinError}}(f^\prime, {\boldsymbol{x}}_{p}),
                        p \in \boldsymbol{P} \}$

            \item Check if the new feature improves the separation: 

            \item[] {\bf if} $\tilde{e}_{min}<{e}_{min}$ {\bf then}
            \begin{algenum}
                \item[] ${e}_{min}=\tilde{e}_{min}$
                \item[] $\boldsymbol{F} \gets \boldsymbol{F} \cup (opt, \alpha)$
                \item[] $\boldsymbol{P} \gets \boldsymbol{P} \setminus opt$
                \item[] update current feature $f^\prime = f^\prime\times\cos(\alpha) + {\boldsymbol{x}}_{opt}\times\sin(\alpha)$
            \end{algenum}
            \item[]{\bf else} \\ \tab {\bf break}

    \end{algenum}

    \item Get the normal vector of the separating hyperplane:
    \begin{algenum}
        \item Declare a vector $\boldsymbol{r} \in \mathbb{R}^d$, $\boldsymbol{r}_i=0$ for all $i=1,...,d$. Set $a = 1$.
        \item Calculate the vector components as
           $\boldsymbol{r}_{{\boldsymbol{F}_i}.p} = \prod_{j=i+1}^{\sharp\boldsymbol{F}} \Bigl(\cos({\boldsymbol{F}_j}.\alpha) \Bigr) \sin({\boldsymbol{F}_i}.\alpha)$:
        \item[] {\bf for all} $i\in \{\sharp\boldsymbol{F}..2\}$ {\bf do}
        \begin{algenum}
            \item[] $\boldsymbol{r}_{{\boldsymbol{F}_i}.p} = a\times\sin({\boldsymbol{F}_i}.\alpha)$
            \item[] $a = a\times\cos({\boldsymbol{F}_i}.\alpha)$
        \end{algenum}
        \item[] $\boldsymbol{r}_{\boldsymbol{F}_1.p} = a$

        \item Project the points on the ray: $\boldsymbol{p}_i.y = y_i$, $\boldsymbol{p}_i.x = \boldsymbol{r}\cdot\tilde{\boldsymbol{x}}_i$
        \item Sort $\boldsymbol{p}$ w.r.t.~$\boldsymbol{p}_{\cdot}.x$  in ascending order.
        \item Count the cardinalities before the separation plane
            \\ $m_{l-} = \sharp\{i:\boldsymbol{p}_i.y = -1, \boldsymbol{p}_i.x\le0\}$, \\$m_{l+} = \sharp\{i:\boldsymbol{p}_i.y = +1, \boldsymbol{p}_i.x\le0\}$
        \item Count the errors
            \\$e_{-} = m_{l+} + m_{-} - m_{l-}$, \\$e_{+} = m_{l-} + m_{+} - m_{l+}$
        \item {\bf if} $e_{-} > e_{+}$ {\bf then}
        \\ \tab $\boldsymbol{r} \gets -\boldsymbol{r}$
    \end{algenum}

\end{enumerate}

Output: the normal vector of the separating hyperplane $r$.

\bigskip

\textbf{Procedure \textsl{GetMinError}}

Input: current feature $f\in \mathbb{R}^n$, property $x \in \mathbb{R}^n$.
\begin{enumerate}
    \item Obtain angles:
    \begin{algenum}
        \item Calculate $\alpha_i = \arctan\frac{{x}_{i}}{{f}_{j}}$, $i=1,...,n$, with $\arctan\frac{0}{0}=0$.
        \item Aggregate angles into set $\mathcal{A}$. Denote $\mathcal{A}_i.\alpha=\alpha_i$ and $\mathcal{A}_i.y=y_i$ the angle and the pattern of the corresponding point. Set $\mathcal{A}_i.y$ to 0 for the points having both $x_i=0$ and $f_i=0$.
        \item Sort $\mathcal{A}$ w.r.t.~$\mathcal{A}_{\cdot}.\alpha$ in ascending order.
    \end{algenum}

    \item Look for the optimal threshold:
    \begin{algenum}
        \item Define $i_{opt} = \arg\max_i \left(|\sum_1^{i}\mathcal{A}_i.y| + |\sum_{i+1}^{n}\mathcal{A}_i.y| \right)$ as the place of the optimal threshold
                   and $e_{min} = n-\max_i \left(|\sum_1^{i}\mathcal{A}_i.y| + |\sum_{i+1}^{n}\mathcal{A}_i.y| \right)$ as the minimal number of incorrectly classified points
        \item Define the optimal angle $\alpha_{opt} = \frac{1}{2}(\mathcal{A}_{i_{opt}+1}.\alpha + \mathcal{A}_{i_{opt}+2}.\alpha) - \frac{\pi}{2}$.
    \end{algenum}
\end{enumerate}

Output: min error $e_{min}$, optimal angle $\alpha_{opt}$

\bigskip

\input{table_alpha} 

From the practical point of view, the routine $DD\alpha$-separator has high computation speed as in each plane it has the complexity of the quick-sort procedure: $O\bigl(\sum_{i=1}^q n_i \log(\sum_{i=1}^q n_i)\bigr)$.

While minimizing empirical risk in two-dimensional coordinate subspaces and due to the choice of efficient for classification features, the $DD\alpha$-separator tends to be \emph{close to} the optimal \emph{risk-minimizing} hyperplane in the extended space.
To a large extent, this explains the performance of the $DD\alpha$-procedure on finite samples.

The robustness of the procedure is twofold: First, \emph{regarding points}, as the depth-space is compact, the outlyingness of the points in it is restricted, and the $DD\alpha$-separator is robust due to its risk-minimizing nature, \textit{i.e.}~by the discrete (zero-or-one) loss function. And second, \emph{regarding features}, the separator is not entirely driven by the exact points' location, but accounts for importance of features of the (extended) depth space. By that, the model complexity is kept low; in practice a few features are selected only, see, \textit{e.g.}, Section~5.2 of~\cite{MozharovskyiML15}.

For theoretical results on the $DD\alpha$-procedure the reader is referred to Section~4 of~\cite{LangeMM14a}. \cite{MozharovskyiML15} provide an extensive comparative empirical study of its performance with a variety of data sets and for different depth notions and outsider treatments, while \cite{LangeMM14b} conduct a simulation study on asymmetric and heavy-tailed distributions.

\subsection[Alternative separators in the $DD$-plot]{Alternative separators in the $DD$-plot}\label{ssec:ddclassAlternatives}
Besides the $DD\alpha$-separator, the package \pkg{ddalpha} allows for two alternative separators in the depth space: a polynomial rule and the $k$-nearest-neighbor ($k$NN) procedure.

When \cite{LiCAL12} introduce the $DD$-classifier, they suggest to use a polynomial of certain degree passing through the origin of the $DD$-plot to separate the two training classes. Based on the fact that by choosing the polynomial order appropriately the empirical risk can be approximated arbitrarily well, they prove the consistency of the $DD$-classifier for a wide range of distributions including some important cases of the elliptically symmetric family. In practice, the minimal error is searched by smoothing the empirical loss with a logistic function and then optimizing the parameter of this function. This strategy has sources of instability such as choice of the smoothing constant and multimodality of the loss function. The authors (partially) solve the last issue by varying the starting point for optimization and multiply running the entire procedure, which increases computation time. For theoretical derivations and implementation details see Sections~4 and~5 of \cite{LiCAL12}. For a simulation comparison of the polynomial rule in the $DD$-plot and the $DD\alpha$-separator see Section~5 of \cite{LangeMM14a}.

In his PhD-thesis, \cite{Vencalek11} suggests to perform the $k$NN classification in the depth space, and proves its consistency for elliptically distributed classes with identical radial densities. For theoretical details and a simulation study see Sections~3.4.3 and~3.7 of \cite{Vencalek11}, respectively. It is worth to notice that the $k$NN-separator has another advantage --- it is directly extendable to more than two classes.


\section[Outsiders]{Outsiders}\label{sec:outsiders}

For a number of depth notions like halfspace, zonoid, or simplicial depth, the depth of a point vanishes beyond the convex hull of the data.
This leads to the problem that new points (to be classified) lying beyond the convex hull of each of the training classes have depth zero w.r.t.~all of them. By that, they are depth-mapped to the origin of the $DD$-plot, and thus cannot be readily classified. We call these points \emph{outsiders} \citep{LangeMM14a}.

\begin{figure}[!h]
  \begin{center}
    \includegraphics[keepaspectratio=true,width = \textwnew, trim=4mm 15mm 4mm 25mm, clip]{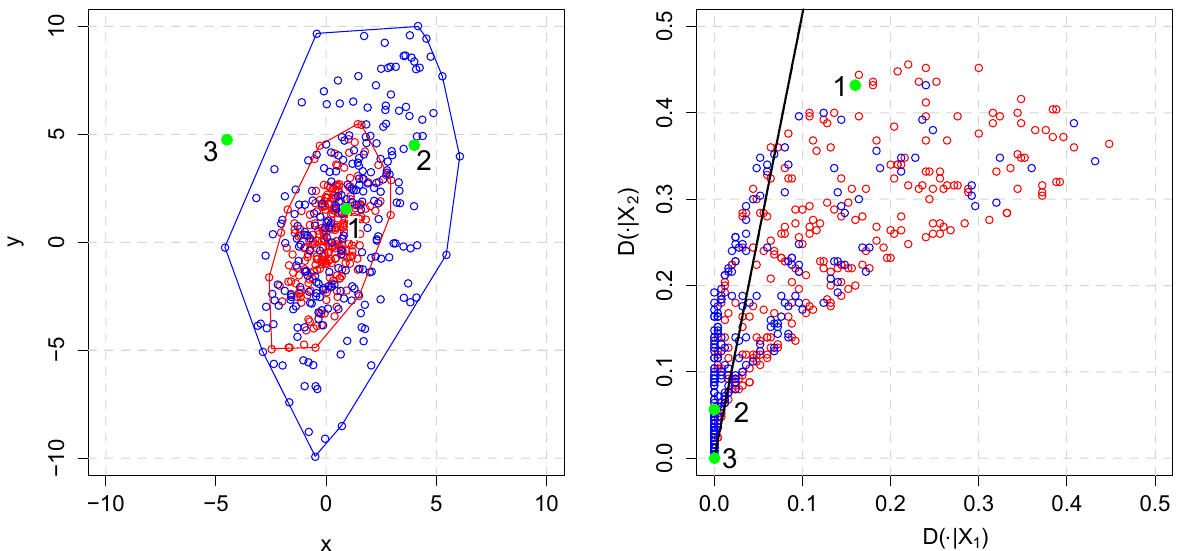}
    \caption{Points to be classified (green) in the original (left) and depth (right) space.}
    \label{fig:outs_demo}
  \end{center}
\end{figure}

Regard Figure~\ref{fig:outs_demo}, where three green points are to be classified. Point ``1'' has positive depth in both classes, and based on its location in the $DD$-plot will be assigned to the less scattered ``red'' class. Point ``2'' has zero depth in the ``red'' class, but a positive one in the more scattered ``blue'' class, to which it will be assigned based on the classification rule in the $DD$-plot. Point ``3'' on the other hand has zero depth w.r.t.~both training classes, and thus classification rule in the $DD$-plot is helpless. Nevertheless, visually it clearly belongs to the ``blue'' class, and most probably would be correctly classified by a very simple classifier, say a poorly tuned $k$NN (\textit{e.g.} 1NN). The suggestion thus is to apply an additional fast classifier to the outsiders.

\begin{figure}[!h]
  \begin{center}
    \includegraphics[keepaspectratio=true,width = \textwnew, trim = 3mm 2mm 0mm 2mm, clip, page = 2]{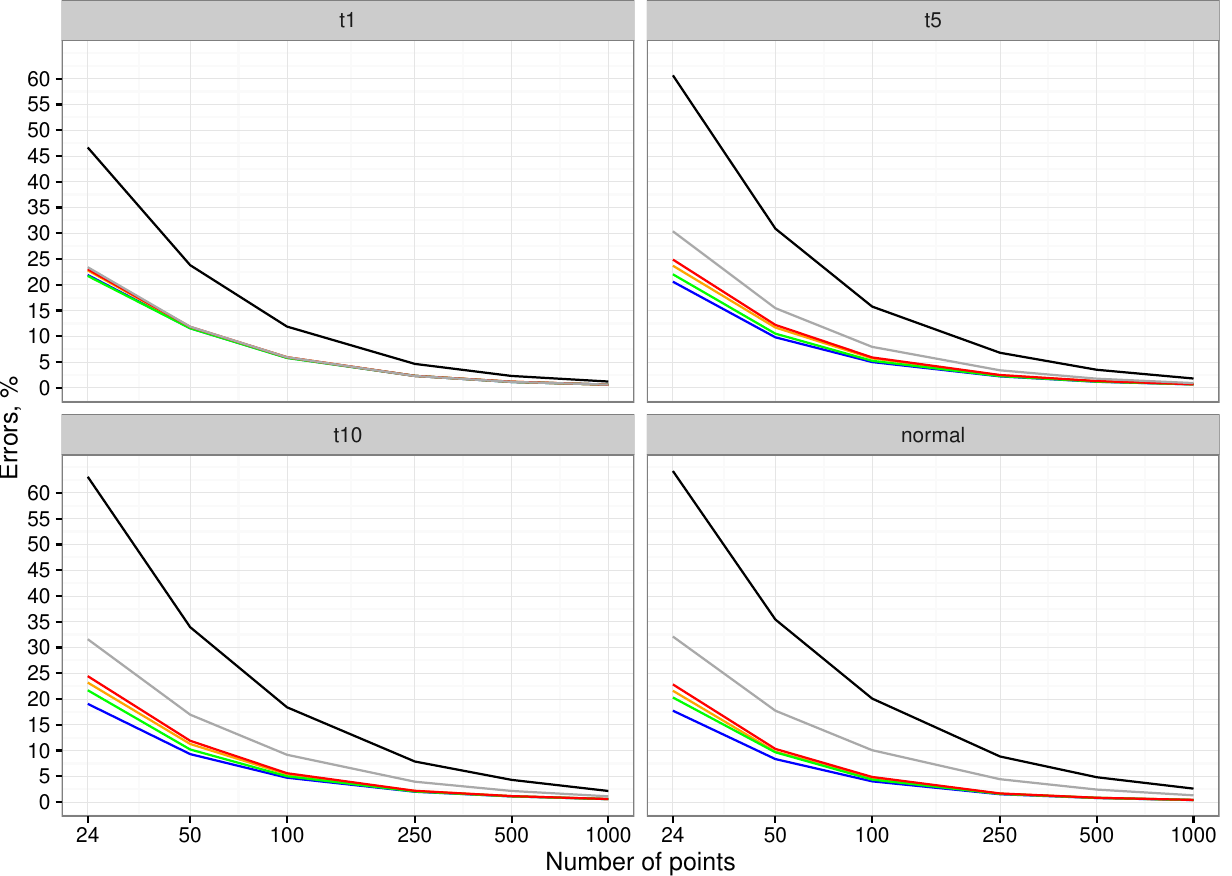}\\
    {\small{\color{black}\LBL} number of outsiders, {\color[gray]{0.5}\LBL} random, {\color{blue}\LBL} LDA, {\color[rgb]{1,0.5,0}\LBL} QDA, {\color{green}\LBL} k-NN, {\color{red}\LBL} Mahalanobis max depth}
    \caption{Error rates of various outsiders treatment. Only outsiders are classified.}
  \label{fig:outs_graphs}
  \end{center}
\end{figure}

The \proglang{R}-package \pkg{ddalpha} implements a number of outsider treatments: linear (LDA) and quadratic (QDA) discriminant analysis, $k$NN, maximum depth classifier based on Mahalanobis depth; and additionally random classification or identification of outsiders for statistical analysis or passing to another procedure. For the same experimental setting as in Section~\ref{ssec:depthMaxdepth}, we contrast these treatments in Figure~\ref{fig:outs_graphs}, comparing classification errors on outsiders only. One can see that for the heavy-tailed Cauchy distribution, where classes may be rather mixed, no outsider treatment performs significantly better than random assignment. The situation improves with increasing number of degrees of freedom of the Student-$t$ distribution, with LDA forming the classification error frontier, as the classes differ in location only. On the other hand, with increasing $n_i$, difference between the treatment becomes negligible. For an extensive comparative study of different outsider treatments the reader is referred to \cite{MozharovskyiML15}.




If outsiders pose a serious problem, one can go for a nowhere-vanishing depth. But in general, the property of generating outsiders should not necessarily be seen as a shortfall, as it allows for additional information when assessing the configured classifier or a data point to be classified. If too many points are identified as outsiders (what can be checked by a validation procedure), this may point onto inappropriate tuning. On the other hand, if outsiders appear extremely rarely in the classification phase (or, \textit{e.g.}, during online learning), an outsider may be an atypical observation not fitting to the data topology in which case one may not want to classify it at all but rather label indicatively.

\section[An extension to functional data]{An extension to functional data}\label{sec:functions}
Similar to Section~\ref{sec:ddclass}, consider a binary classification problem in the space of real valued functions defined on a compact interval, which are continuous and smooth everywhere except for a finite number of points, \textit{i.e.}~given two classes of functions: $\boldsymbol{\mathcal{F}}_1=\{f_1,...,f_{n_1}\}$ and $\boldsymbol{\mathcal{F}}_2=\{f_1,...,f_{n_2}\}$, again indexing observations by $i=1,...,n_1,n_1 + 1,...,n_1 + n_2$ for convenience. (An aggregation scheme extends this binary classification to the multiple one.) The natural extension of the depth-based classification to the functional setting consists in defining a proper depth transform $(\boldsymbol{\mathcal{F}}_1,\boldsymbol{\mathcal{F}}_2)\mapsto\boldsymbol{Z}=\{\boldsymbol{z}_i=(D(f_i|\boldsymbol{\mathcal{F}}_1),D(f_i|\boldsymbol{\mathcal{F}}_2)),\,i=1,...,n_1+n_2\}$ similar to that in Section~\ref{sec:ddclass}. For this, a proper functional depth should be employed \citep[see][and references therein for an overview]{MoslerP12,NietoReyesB16}, followed by the suitable classification technique in the (finite dimensional) depth space. As the functional data depth reduces space dimensionality from infinity to one, the final performance is sensitive to the choice of the depth representation and of the finite-dimensional separator, and thus both constituents should be chosen very carefully. Potentially, this lacks quantitative flexibility because of the finite set of existing components. Nevertheless, in many cases this solution provides satisfactory results; see a comprehensive discussion by \cite{CuestaAlbertosFBOdlF15} with experimental comparisons involving a number of functional depth notions and $q$-dimensional classifiers, as well as their implementation in the \proglang{R}-package \pkg{fda.usc}. Corresponding functional depth procedures can also be used with R-package \pkg{ddalpha}, see Section~\ref{sec:application} for a detailed explanation.

\pkg{ddalpha} suggests two implementations of the strategy of immediate functional data projection onto a finite-dimensional space with further application of a multivariate depth-based classifier: componentwise classification by \cite{DelaigleHB12} and $LS$-transform proposed by \cite{MoslerM15}. Both methodologies allow to control for the quality of classification in a quantitative way (\textit{i.e.}~by tuning parameters) when constructing the multivariate space, which in addition enables consistency derivations. For the first one the reader is referred to the literature; the second one we present right below.

In application, functional data is usually given in a form of discretely observed paths $\boldsymbol{\tilde f}_i = \left[f_{i}(t_{i1}), f_{i}(t_{i2}), ..., f_{i}(t_{iN_i})\right]$, which are the measurements at ordered (time) points $t_{i1}< t_{i2}< ... <t_{iN_i}$, $i=1,...,n_1+n_2$, not necessarily equidistant nor same for all $i$. Fitting these to a basis is avoided as the choice of such a basis turns out to be crucial for classification and thus should better not be independently selected prior to it. Instead, a simple scheme is suggested based on integrating linearly extrapolated data and their derivatives over a chosen number of intervals. Let $\min_i{t_{i1}}=0$ and let $T=\max_i{t_{iN_i}}$, then one obtains the following finite-dimensional transform:
{\footnotesize
\begin{equation}\label{equ:average}
{\hat f}_{i}\mapsto\boldsymbol{x}_{i} = \Bigl[\int_0^{T/L} {\hat f}_{i}(t) dt, \dots,  \int_{T(L-1)/L}^{T} {\hat f}_{i}(t) dt, \int_0^{T/S} {\hat f'}_{i}(t) dt, \dots, \int_{T(S-1)/S}^{T} {\hat f'}_{i}(t) dt \Bigr]\,,
\end{equation}
}
with ${\hat f}_{i}(t)$ being the function obtained by connecting the points $(t_{ij},\, f_{i}(t_{ij})), j=1,\dots, N_i$ with line segments and setting ${\hat f}_{i}(t)=f_{i}(t_{i1})$ when $0\le t\le t_{i1}$ and ${\hat f}_{i}(t)=f_{i}(t_{ik_i})$ when $t_{iN_i}\le t\le T$, ${\hat f'}_{i}(t)$ being its derivative, and $L,S\ge 0$, $L+S\ge 2$ being integers. $L$ and $S$ are the numbers of intervals of equivalent length to integrate over the location and the slope of the function, and have to be tuned. One can use intervals of different length or take into account higher-order derivatives (constructed as differences, say), but the suggested way appears to be simple and flexible enough. Moreover it does not introduce any spurious information. The set of considered $LS$-pairs can be chosen on the basis of some prior knowledge about the nature of the functions or just by properly restricting the dimension of the constructed space by $d_{min}\le L + S \le d_{max}$. Cross-validation is then used to choose the best $LS$-pair. \pkg{ddalpha} suggests to reduce the set of cross-validated $LS$-pairs by employing the Vapnik-Chervonenkis bound. The idea behind is that, while being conservative, the bound can still provide insightful ordering of the $LS$-pairs, especially in the case when the empirical risk and the bound have the same order of magnitude.

Given a set of considerable pairs $\mathcal{S}=\{(l_i,s_i)|i=1,...,N_{ls}\}$, for each its element calculate the Vapnik-Chervonenkis bound \citep[see][for this particular derivation]{MoslerM15}
\begin{equation}\label{equ:Vapnik}
  b^{VC}_i = \epsilon\left(\boldsymbol{c},\boldsymbol{\mathcal{\hat F}}_1^{(l_i,s_i)},\boldsymbol{\mathcal{\hat F}}_2^{(l_i,s_i)}\right) + \sqrt{\frac{\ln{2\sum_{k=0}^{l_i + s_i - 1}{n_1 + n_2 - 1 \choose k}} - \ln{\eta}}{2(n_1 + n_2)}},
\end{equation}
where $\epsilon\left(\boldsymbol{c},\boldsymbol{\mathcal{\hat F}}_1^{(l_i,s_i)},\boldsymbol{\mathcal{\hat F}}_2^{(l_i,s_i)}\right)$ is the empirical risk achieved by a linear classifier $\boldsymbol{c}$ on the data transformed according to (\ref{equ:average}) with $L=l_i$, $S=s_i$ and $1-\eta$ is the chosen reliability level. In \pkg{ddalpha} we set $\eta=\frac{1}{n_1 + n_2}$, and choose $\boldsymbol{c}$ to be the LDA for its simplicity and speed.
Then a subset $\mathcal{S}^{CV}\subset \mathcal{S}$ is chosen possessing the smallest values of $b^{VC}_i$: $\bigl((l_j,s_j)\in \mathcal{S}^{CV}, (l_k,s_k)\in \mathcal{S}\setminus \mathcal{S}^{CV}\bigr)$ $\Rightarrow$ $(b^{VC}_j < b^{VC}_k)$, and cross-validation is performed over all $(l,s)\in \mathcal{S}^{CV}$. For the subsample referenced in introduction, the functions' levels and slopes are shown in Figure~\ref{fig:funcdata}; the $LS$-representation is selected by reduced cross-validation due to (\ref{equ:average}) having $(L,S)=(1,1)$, and is depicted in Figure~\ref{fig:discretespace}, left.

\section{Usage of the package}\label{sec:application}

\begin{figure}[!h]
  \begin{center}
  \includegraphics[keepaspectratio=true,width = \textwnew, trim = 30mm 8mm 57mm 61mm, clip, page = 1]{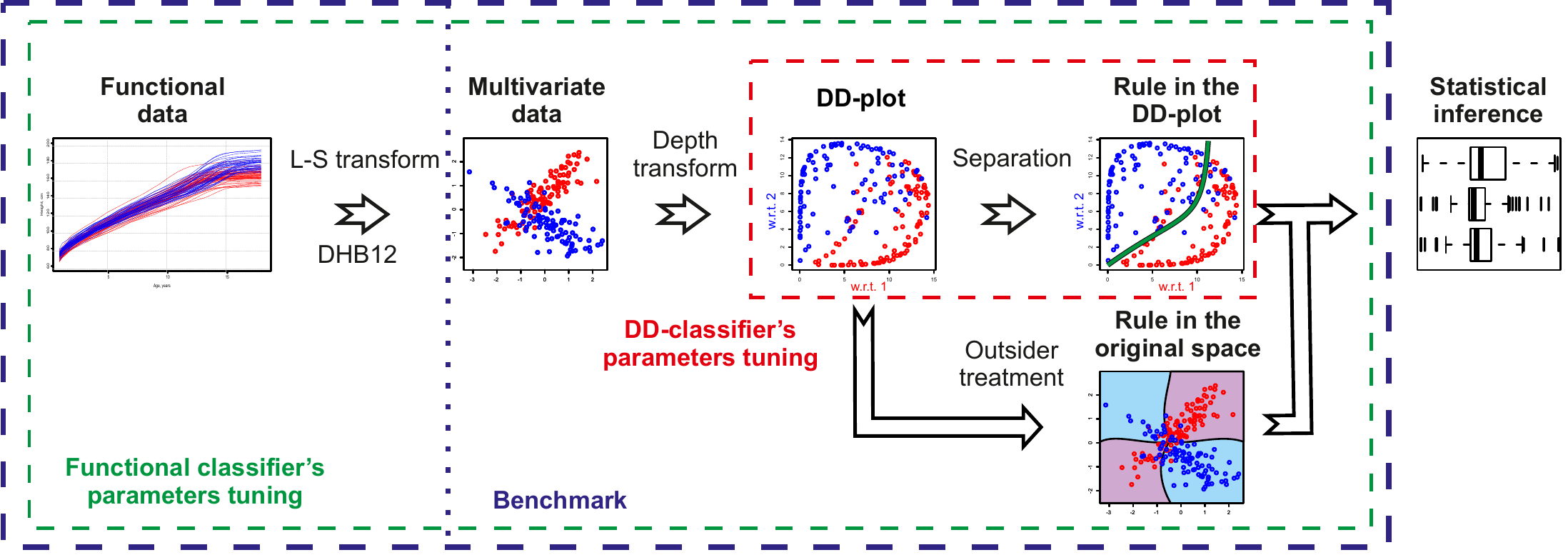}
  \caption{The structure of the package}
  \label{fig:ddalpha_structure}
  \end{center}
\end{figure}

The package \pkg{ddalpha} is a structured solution that provides computational machinery for a number of depth functions and classifiers for multivariate and functional data. It also allows for user-defined depth functions and separators in the $DD$-plot (further $DD$-separators). The structure of the package is presented in Figure~\ref{fig:ddalpha_structure}.

\subsection{Basic functionality}

Primary aims of the package are calculation of data depth and depth-classification.

\textbf{\emph{Data depth}} is calculated by calling
\begin{Code}
depth.<depthName>(x, data, ...),
\end{Code}
where \code{data} is a matrix with each row being a $d$-variate point, and \code{x} is a matrix of objects whose depth is to be calculated. Additional arguments (\code{...}) differ between depth notions. The output of the function is a vector of depths of points from \code{x}.
Most of the depth functions possess both exact and approximative versions that are toggled with parameters \code{exact} and \code{method}, see Table~\ref{tab:depths_impl}.
The exact algorithms of Mahalanobis, spatial, and zonoid depths are very fast and thus exclude the need of approximation.
Mahalanobis and spatial depths use either traditional moment or MCD estimates of mean and covariance matrix.
Methods \code{random} for projection depth and \code{Sunif.1D} for halfspace depth approximate the depth as the minimum univariate depth of the data projected on \code{num.directions} directions uniformly distributed on $S^{d-1}$. The exact algorithms for the halfspace depth implement the framework described in Section~\ref{ssec:depthNotions}, where the dimensionality $k$ of the combinatorial space is specified as follows: $k=1$ for method \code{recursive}, $k=d-2$ for \code{plane} and $k=d-1$ for \code{line}, see additionally~\cite{DyckerhoffM16}.
The second approximating algorithm for projection depth is \code{linearize} --- the Nelder-Mead method for function minimization, taken from \cite{NelderM65} and originally implemented in \proglang{R} by Subhajit Dutta.
For simplicial and simplicial volume depths, parameter \code{k} specifies the number (if \code{k} $>1$) or portion (if $0<$ \code{k} $<1$) of simplices chosen randomly among all possible simplices for approximation.

\newpage 
\begin{table}[!th]
\centering
{\small
\begin{tabular}{llll}
\hline
Depth & Exact & Approximative & Parameter\\
\hline
Mahalanobis & \code{moment} &   & \code{mah.estimate}\\
            & \code{MCD}   & & \\
spatial     & \code{moment}  & & \code{mah.estimate}\\
            & \code{MCD}   & & \\
            & \code{none}   & & \\
projection  &               & \code{random} &  \code{method}\\
            &               & \code{linearize} & \\
halfspace   & \code{recursive} & \code{Sunif.1D} & \code{method}\\
            & \code{plane}  & & \\
            & \code{line}   & & \\
simplicial  & + & + & \code{exact}\\
simplicial volume& + & + & \code{exact}\\
zonoid      & + & \\
\hline
\end{tabular}}
\caption{Implemented depth algorithms}\label{tab:depths_impl}
\end{table}

In addition, calculation of the entire $DD$-plot at once is possible by
\begin{Code}
depth.space.<depthName>(data, cardinalities, ...),
\end{Code}
where the matrix \code{data} consists of $q$ stacked training classes, and \code{cardinalities} is a vector containing numbers of objects in each class. The method returns a matrix with $q$ columns representing the depths of each point w.r.t.~each class.

\textbf{\emph{Classification}} can be performed either in two steps --- training the classifier with \code{ddalpha.train} and using it for classification in \code{ddalpha.classify}, or in one step --- by function \code{ddalpha.test(learn, test, ...)} that trains the classifier with \code{learn} sample and checks it on the \code{test} one. Other parameters are the same as for function \code{ddalpha.train} and are described right below.

Function \code{ddalpha.train} is the main function of the package. Its structure is shown on the right part of Figure~\ref{fig:ddalpha_structure}.
\begin{Code}
ddalpha.train(data, depth = "halfspace", separator = "alpha",
  outsider.methods = "LDA", outsider.settings = NULL,
  aggregation.method = "majority",
  use.convex = FALSE,
  seed = 0, ...)
\end{Code}
The notion of the depth function and the $DD$-separator are specified with the parameters \code{depth} and \code{separator}, respectively.
Parameter \code{aggregation.method} determines the method applied to aggregate outcomes of binary classifiers during multiclass classification. When \code{"majority"}, $q(q-1)/2$ binary one-against-one classifiers are trained, and for \code{"sequent"}, $q$ binary one-against-all classifiers are taught. During classification, the results are aggregated using the majority voting, where classes with larger proportions in the training sample are preferred when tied (by that implementing both aggregating schemes at once).
Additional parameters of the chosen depth function and $DD$-separator are passed using the dots, and are described in the help sections of the corresponding \proglang{R}-functions.
Also, the function allows to use a pre-calculated $DD$-plot by choosing \code{depth = "ddplot"}.
For each depth function and depth-separator, a validator is implemented --- a special \proglang{R}-function that specifies the default values and checks the received parameters allowing by that definition of custom depths and separators; see Section~\ref{ssec:custom_dsss} for details.

\textbf{\emph{Outsider treatment}} is a supplementary classifier for data that lie outside the convex hulls of all $q$ training classes. It is only needed during classification when the used data depth produces outsiders or obtains zero values in the neighborhood of the data.
Parameter \code{use.convex} of \code{ddalpha.train} indicates whether outsiders should be determined as the points not contained in any of the convex hulls of the classes from the training sample (\code{TRUE}) or those having zero depth w.r.t.~each class from the training sample (\code{FALSE}); the difference is explained by the depth approximation error.
The following methods are available:
\code{"LDA"}, \code{"QDA"} and \code{"kNN"}; affine-invariant $k$NN (\code{"kNNAff"}), \textit{i.e.}~$k$NN with Euclidean distance normalized by the pooled covariance matrix, suited only for binary classification and using aggregation with multiple classes and not accounting for ties, but very fast; maximum Mahalanobis depth classifier (\code{"depth.Mahalanobis"}); equal and proportional randomization (\code{"RandEqual"} and \code{"RandProp"}) and ignoring (\code{"Ignore"}) --- a string ``Ignored'' is returned for the outsiders.
Outsider treatment is set by means of parameters \code{outsider.methods} and \code{outsider.settings} in \code{ddalpha.train}. Multiple methods may be trained and then the particular method is selected in \code{ddalpha.classify} by passing its name to
parameter \code{outsider.method}.
Parameter \code{outsider.methods} of \code{ddalpha.train} accepts a vector of names of basic outsider methods that are applied with the default settings.
Parameter \code{outsider.settings} allows to train a list of outsider treatments, whose elements specify the names of the methods (used in \code{ddalpha.classify} later) and their parameters.

\textbf{\emph{Functional classification}} is performed with functions \code{ddalphaf.train} implementing $LS$-transform \citep{MoslerM15} and \code{compclassf.train} implementing componentwise classification \citep{DelaigleHB12}.
\begin{Code}
ddalphaf.train(dataf, labels,
  adc.args = list(instance = "avr",
    numFcn = -1,
    numDer = -1),
  classifier.type = c("ddalpha", "maxdepth",
    "knnaff", "lda", "qda"),
  cv.complete = FALSE,
  maxNumIntervals = min(25, ceiling(length(dataf[[1]]$args)/2)),                                        $ closing dollar, MUST NOT BE PRINTED
  seed = 0,
  ...)
\end{Code}
\begin{Code}
compclassf.train(dataf, labels,
  to.equalize = TRUE,
  to.reduce = TRUE,
  classifier.type = c("ddalpha", "maxdepth",
    "knnaff", "lda", "qda"),
  ...)
\end{Code}

In both functions, \code{dataf} is a list of functional observations, each having two vectors: \code{"args"} for arguments sorted in ascending order and \code{"vals"} for the corresponding functional evaluations;
\code{labels} is a list of class labels of the functional observations;
\code{classifier.type} selects the classifier that separates the finitized data, and additional parameters are passed to this selected classifier with dots.
In the componentwise classification, \code{to.equalize} specifies whether the data is adjusted to have equal (the largest) argument interval, and \code{to.reduce} indicates whether the data has to be projected onto a low-dimensional space via the principal components analysis (PCA) in case its affine dimension after finitization is lower than expected.
(Both parameters are recommended to be set true.)

The $LS$-transform converts functional data into multidimensional ones by averaging over intervals or evaluating values on equally-spaced grid for each function and its derivative on $L$ 
(respectively $S$) equal nonoverlapping covering intervals.
The dimension of the multivariate space then equals $L+S$.
Parameter \code{adc.args} is a list that specifies: \code{instance} --- the type of discretization of the functions having values \code{"avr"} for averaging over intervals of the same length
and \code{"val"} for taking values on equally-spaced grid;
\code{numFcn} ($L$) is the number of function intervals, and \code{numDer} ($S$) is the number of first-derivative intervals.

The parameters $L$ and $S$ may be set explicitly or may be automatically cross-validated. The cross-validation is turned on by setting \code{numFcn = -1} and \code{numDer = -1}, or by passing a list of \emph{adc.args} objects to \code{adc.args} --- the range of $(L,S)$-pairs to be checked. In the first case all possible pairs of $L$ and $S$ are considered up to the maximal dimension that is set in \code{maxNumIntervals}, while in the latter case only the pairs from the list are considered. The parameter \code{cv.complete} toggles the complete cross-validation; if \code{cv.complete} is set to false the Vapnik-Chervonenkis bound is applied, which enormously accelerates the cross-validation, as described in \cite{MoslerM15} in detail. 
The optimal values of $L$ and $S$ are stored in the \emph{ddalphaf} object, that is returned from \code{ddalphaf.train}.

\subsection{Custom depths and separators}\label{ssec:custom_dsss}

As mentioned above, the user can amplify the existing variety by defining his own depth functions and separators. Custom depth functions and separators are defined by implementing three functions: parameters validator, learning, and calculating functions, see Tables~\ref{tab:custom_depth} and~\ref{tab:custom_separator}.
Usage examples are found in the manual of the package \pkg{ddalpha}.

\input{tables_custom} 

\emph{Validator} is a nonmandatory function that validates the input parameters and checks if the depth calculating procedure is applicable to the data. All the parameters of a user-defined depth or separator must be returned by a validator as a named list, otherwise they will not be saved in the \code{ddalpha} object.

\textbf{\emph{Definition of a custom depth function}} is done as follows:
The \emph{depth-training function} \code{.<name>_learn(ddalpha)} 
calculates any data-based statistics that the depth function needs (\textit{e.g.}, mean and covariance matrix for Mahalanobis depth) and then calculates the depths of the training classes, \textit{e.g.}, by calling for each pattern $i$ the \emph{depth-calculating function} \code{.<name>_depths(ddalpha, objects = ddalpha$patterns[[i]]$points)} that calculates the depth of each point in \code{objects} w.r.t.~each pattern in \code{ddalpha} and returns a matrix with $q$ columns.
The learning function returns a \code{ddalpha} object, where the calculated statistics and parameters are stored. All stored objects, including the parameters returned by the validator, are accessible through the \code{ddalpha} object, on each stage.
After having defined these functions, the user only has to specify \code{depth = "<name>"} in \code{ddalpha.train} and pass the required parameters there. (The functions are then linked via the \code{match.fun} method.)

\textbf{\emph{Definition of a custom separator}} is similar. Recall that there exist binary separators applicable to two classes, and multiclass ones that separate more than two classes at once. In case if the custom method is binary, the package takes care of the voting procedures, and the user only has to implement a method that separates two classes.
The training method for a \emph{binary separator} \code{.<name>_learn(ddalpha, index1, index2, depths1, depths2)} accepts the depths of the objects w.r.t.~two classes and returns a trained classifier.
A \emph{multiclass separator} has to implement another interface: \code{.<name>_learn(ddalpha)}, accessing the depths of the different classes via \code{ddalpha$patterns[[i]]$depths}.
The binary classifier can utilize the whole depth space (\textit{i.e.}~depths w.r.t.~other classes than the two currently under consideration) to get more information like the $\alpha$-separator does, or restrict to the $DD$-plot w.r.t.~the two given classes like the polynomial separator, by accessing \code{depths1} and \code{depths2} matrices.
The \emph{classifying function} \code{.<name>_classify(ddalpha, classifier, objects)} accepts the previously trained \code{classifier} and the depths of the objects that are classified.
For a binary classifier, the indices of the currently classified patterns are accessible as \code{classifier$index1} and \code{classifier$index2}.
A binary classifier shall return a vector with positive values for the objects from the first class, and the multiclass classifier shall assign to each object to be classified the index of the corresponding pattern in \code{ddalpha}.
Similarly to the depth function, the defined separator is accessible by \code{ddalpha.train} by specifying \code{separator = "<name>"}. If a nonbinary method is used, it is important to set \code{aggregation.method = "none"} or (preferred but more complicated) to return \code{ddalpha$methodSeparatorBinary = F} from the validator, otherwise the method will be treated as a binary one, as by default \code{aggregation.method = "majority"}.

\subsection{Additional features}\label{ssec:addFeatures}

A number of additional functions are implemented in the package to facilitate assessing quality and time of classification, handle multimodally distributed classes, and visualize depth statistics.

\textbf{\emph{Benchmark procedures}} implemented in the package allow for estimating expected error rate and training time:
\begin{Code}
ddalpha.test(learn, test, ...)
ddalpha.getErrorRateCV(data, numchunks = 10,  ...)
ddalpha.getErrorRatePart(data, size = 0.3, times = 10,  ...)
\end{Code}
The first function trains the classifier on the \code{learn} sample, checks it on the \code{test} one, and reports the error rate, the training time and other related values such as the numbers of correctly and incorrectly classified points, number of ignored outsiders, \textit{etc}.
The second function performs a cross-validation procedure over the given data. On each step, every \code{numchunks}\textit{th} observation is removed from the data, the classifier is trained on these data and tested on the removed observations. The procedure is performed until all points are used for testing. Setting \code{numchunks} to $n$ leads to the leave-one-out cross-validation (=jackknife) that is a consistent estimate of the expected error rate. The procedure returns the error rate, \textit{i.e.}~the total number of incorrectly classified objects divided by the total number of objects.
The third function performs a benchmark procedure by partitioning the given data. On each of \code{times} steps, randomly picked \code{size} observations are removed from the data, the classifier is trained on these data and tested on the removed observations.
The outputs of this function are the vector of errors, their mean and standard deviation. Additionally, both functions report mean training time and its standard deviation.
In all three functions, dots denote the additional parameters passed to \code{ddalpha.train}.
Benchmark procedures may be used to \emph{tune the classifier} by setting different values and assessing the error rate.
The function \code{ddalpha.test} is more appropriate for simulated data, while the two others are more suitable for subsampling learning with real data  and testing sequences from it.
Analogs of these procedures for a functional setting are present in the package as well:
\begin{Code}
ddalphaf.test(learn, learnlabels, test, testlabels, disc.type, ...)
ddalphaf.getErrorRateCV(dataf, labels, numchunks, disc.type,  ...)
ddalphaf.getErrorRatePart(dataf, labels, size, times, disc.type,  ...)
\end{Code}
The discretization scheme is chosen with parameter \code{disc.type} setting it to \code{"LS"} or \code{"comp"}.
Note that these procedures are made to assess the error rates and the learning time for a single set of parameters. If the $LS$-transform is used, the parameters $L$ and $S$ shall be explicitly set with \code{adc.args} rather then cross-validated.

\textbf{\emph{Several approaches reflecting multimodality}} of the underlying distribution are implemented in the package.
These methods appear to be useful if the data substantially deviate from elliptical symmetry (\textit{e.g.} having nonconvex or nonconnected support) and the classification based on a global depth fails to achieve close to optimal error rates.
The methods need more complicated and fine parameter tuning, whose detailed description we leave to the corresponding articles.

{\emph{Localized spatial depth}} and a classifier based on it, proposed by \cite{DuttaG15}, can be seen as a $DD$-classifier.
The global spatial depth calculates the average of the unit vectors pointing from the points from $\boldsymbol{X}$ in direction $\boldsymbol{z}$. We rewrite (\ref{equ:sptDepth}) denoting $\boldsymbol{t}_i = \boldsymbol{\Sigma}^{-\frac{1}{2}}(\boldsymbol{X})(\boldsymbol{z} - \boldsymbol{x}_i$)
$$   D_{spt}(\boldsymbol{z}|\boldsymbol{X})=1-\Bigl\|\frac{1}{n}\sum_{i=1}^n \boldsymbol{v}\bigl(\boldsymbol{t}_i\bigr)\Bigr\|.$$
The local version is obtained by kernelizing the distances
$$D_{Lspt}(\boldsymbol{z}|\boldsymbol{X})=\Bigl\|\frac{1}{n}\sum_{i=1}^n K_h(\boldsymbol{t}_i)\Bigr\|-
\Bigl\|\frac{1}{n}\sum_{i=1}^n K_h(\boldsymbol{t}_i) \boldsymbol{v}(\boldsymbol{t}_i)\Bigr\|,$$
with the Gaussian kernel function $K_h(\boldsymbol{x})$.
The bandwidth parameter $h$ defines the localization rate.
(If $h>1$, the depth is multiplied by $h^d$.)

{\emph{The potential-potential (pot-pot) plot}} \citep{PokotyloM16} bears the analogy to the $DD$-plot and thus can be directly used in $DD$-classification as well.
The potential of a class $j$ is defined as a kernel density estimate multiplied by the class's prior probability and is used in the same way as a depth
$$\hat\phi_j(\boldsymbol{x}) = p_j \hat f_j(\boldsymbol{x}) = \frac{1}{n} \sum_{i=1}^{n_j}{K_{\boldsymbol{H}_j}(\boldsymbol{x},\boldsymbol{x}_{ji})},$$
with a Gaussian kernel $K_{\boldsymbol{H}}(\boldsymbol{x})$ and bandwidth matrix $\boldsymbol{H} = h^2\hat{\boldsymbol{\Sigma}}(\boldsymbol{X})$.
The bandwidth parameter $h$ (called \code{kernel.bandwidth} in the package) is separately tuned for each class.
The parameters have to be properly tuned, using the following benchmark procedures:
\begin{Code}
min_error = list(a = NA, error = 1)
for (h in list(c(h_11, h_21), ... , c(h_1k, h_2k)))
{
  error = ddalpha.getErrorRateCV(data, numchunks = <nc>,
    separator = <sep>, depth = "potential", kernel.bandwidth = h,
    pretransform = "NMahMom")
  if(error < min_error$error)
    min_error = list(a = a, error = error)
}
\end{Code}

{\emph{The depth-based $k$NN}} \citep{PaindaveineVB15} is an affine-invariant version of the $k$-nearest-neighbor procedure. This method is different, in the sense that it is not using the $DD$-plot. It is accessible through functions \code{dknn.train}, \code{dknn.classify} and \code{dknn.classify.trained}. For each point $\boldsymbol{x}_0$ to be classified, data points are appended by their reflection w.r.t.~$\boldsymbol{x}_0$, which results in the extended centrally symmetric data set of size $2n$. Then the depth of each data point is calculated in this extended data cloud, and $\boldsymbol{x}_0$ is assigned to the most representable class among $k$ points with the highest depth value, breaking ties randomly. Each depth notion may be inserted. Training the classifier constitutes in its tuning by the leave-one-out cross-validation. The method is integrated into the benchmark procedures, accessible there by setting \code{separator = "Dknn"}.

\textbf{\emph{Depth visualization}} functions applicable to the two-dimensional data are also implemented in the package.
To visualize a depth function as a three-dimensional landscape, use
\begin{Code}
depth.graph(data, depth_f,
  main, xlim, ylim, zlim, xnum, ynum, theta, phi, bold = F, ...)
\end{Code}
The function accepts additional parameters: plot-limiting parameters \code{xlim}, \code{ylim}, \code{zlim} are calculated automatically, parameters \code{xnum}, \code{ynum} control the resolution of the plot, parameters \code{theta} and \code{phi} rotate the plot, and with parameter \code{bold} equal to \code{TRUE} the data points are drawn in bold face.

Depth contours are pictured by the following functions:
\begin{Code}
depth.contours(data, depth,
  main, xlab, ylab, drawplot = T, frequency=100, levels = 10, col, ...)
depth.contours.ddalpha(ddalpha,
  main, xlab, ylab, drawplot = T, frequency=100, levels = 10)
\end{Code}
Function \code{depth.contours} calculates and draws the depth contours $D_\alpha$ for given \code{data}. Parameter \code{frequency} controls the resolution of the plot, and parameter \code{levels} controls the vector of depth values of $\alpha$ for which the contours are drawn. Note that a single value set as \code{levels} defines either the depth of a single contour ($0< $ \code{levels} $ \le 1$) or the number (as its ceiling) of contours that are equally gridded between zero and maximal depth value (\code{levels} $ >1$). To combine the contours of several data sets or several different depth notions in one plot, parameter \code{drawplot} should be set to \code{FALSE} for all but the first plot and the color should be set individually through \code{col}.
It is also possible to draw depth contours for a previously trained \code{ddalpha} classifier. In this case classes will differ in colors.

Figures~\ref{fig:bivariate_depths1} and~\ref{fig:bivariate_depths2} show depth surface (left) and depth contours (right) for each of the implemented depth notions. The two plots, \textit{e.g.} for Mahalanobis depth, correspond (without additional parameters that orientate the plot) to the calls \code{depth.graph(data, "Mahalanobis")} and \code{depth.contours(data, "Mahalanobis")}.

Another useful function draws the $DD$-plot either from the trained $DD\alpha$-classifier or from the depth space, additionally indicating the separation between the classes:
\begin{Code}
draw.ddplot(ddalpha, depth.space, cardinalities,
  main = "DD plot", xlab = "C1", ylab = "C2",
  classes = c(1, 2), colors = c("red", "blue", "green"), drawsep = T)
\end{Code}

To facilitate saving the default parameters for the plots and resetting them, which may become annoying when done often, function \code{par(resetPar())} can be used.

\textbf{\emph{Multivariate and functional data sets}} and data generators are included in the package \pkg{ddalpha} to make the empirical comparison of different classifiers and data depths easier.
50 real multivariate binary classification problems were gathered and described by \cite{MozharovskyiML15} and are also available at \url{http://www.wisostat.uni-koeln.de/de/forschung/software-und-daten/data-for-classification/}. The data can be loaded to a separate variable with function \code{variable = getdata("<name>")}.
Class labels are in the last column of each data set.
Functional data sets are accessible through functions \code{dataf.<name>()} and contain four functional data sets and two generators from \cite{CuevasFF07}.
A functional data object contains a list of functional observations, each characterized by two vectors of coordinates, the arguments vector \code{args} and the values vector \code{vals}, and a list of class labels.
Although this format is clear, visualization of such data can be a nontrivial task, which is solved by function \code{plotf}.

\subsection{Tuning the classifier}

Classification performance depends on many aspects: chosen depth function, separator, outsider treatment, and their parameters.

When selecting a depth function, such properties as ability to reflect asymmetry and 
shape of the data, robustness, vanishing beyond the convex hull of the data, and computational burden have to be considered.

Depth contours of Mahalanobis depth are elliptically symmetric and those of projection depth are centrally symmetric, thus both are not well suited for skewed data. Contours of spatial depth are also rounded, but fit substantially closer to the data, which can also be said about simplicial volume depth. Being intrinsically nonparametric, halfspace, simplicial, and zonoid depths fit closest to the geometry of the data cloud, but vanish beyond its convex hull, and thus produce outsiders during classification.
All these depths are global and not able to reflect localities possibly present in the data.
Local spatial depth as well as potentials compensate for this by fitting multimodal distributions well, which is bought at the price of computational burden for tuning a parameter due to an application specific criteria.

Halfspace, simplicial, and projection depths are robust, while outlier sensitivity of Mahalanobis and spatial depths depends on the underlying estimate of the covariance matrix.
To obtain their robust versions, the MCD estimator is applied in package \pkg{ddalpha}. Parameter \code{mah.parMcd} used with Mahalanobis and spatial depths corresponds to the portion of the data for which the covariance determinant is minimized. Simplicial volume and zonoid depths, being based on volume and mean, fail to be robust in general as well.

Halfspace, zonoid, and simplicial depths produce outsiders; their depth contours are also not smooth, and the contours of the simplicial depth are even star-shaped. These depths must not be considered if a substantial portion of points lies on the convex hull of the data cloud; in some cases, especially in high dimensions, this may reach 100\%, see also \cite{MozharovskyiML15}.

Most quickly computable are Mahalanobis, spatial, and zonoid depths.
Their  calculation speed depends minorly on data dimension and moderately on the size of the data set, while computation time for simplicial, simplicial volume, and exact halfspace depths dramatically increases with the number of points and dimension of the data.
Approximating algorithms balance between calculation speed and precision depending on their parameters.
Random halfspace and projection depths are driven by parameter \code{num.directions}, \textit{i.e.}~the number of directions used in the approximation.
The approximations of simplicial and simplicial volume depths depend on the number of simplices picked, which is set with parameter \code{k}. If a fixed number of simplices \code{k} $>1$ is given the algorithmic complexity is polynomial in $d$ but is independent of $n$, given \code{k}. If a proportion of simplices is given ($0<$ \code{k} $<1$), then the corresponding portion of all simplices is used and the algorithmic complexity is exponential in $n$, but one can assume that the approximation precision is kept on the same level when $n$ changes.
Note that in $\mathbb{R}^2$, the exact efficient algorithm of \cite{RousseeuwR96} is used to calculate simplicial depth.

Based on the empirical study using real data \citep{PokotyloM16}, the classifiers' error rates grow in the following order: $DD\alpha$, polynomial classifier, $k$NN; although $DD\alpha$ and the polynomial classifier provide similar polynomial solutions and $k$NN sometimes delivers good results when the other two fail.
The degree of the $DD\alpha$ and the polynomial classifier and the number of nearest neighbors are automatically cross-validated, but maximal values may be set manually. To gain more insights, depth-transformed data may be plotted (using \code{draw.ddplot}).

The outsider treatment should not be regarded as the one that gives the best separation of the classes in the original space, but rather be seen as a computationally cheap solution for points right beyond their convex hulls.

In functional classification, parameters $L$ and $S$ can be set by the experience-guided applicant or determined automatically by means of cross-validation. The ranges for cross-validation can be based on previous knowledge of the area or conservatively calculated.

Benchmark procedures that we included in the package may be used for empirical parameters' tuning, by iterating the parameters values and estimating the error rates. For example, the following code fragment searches for the separator, depth, and some other parameters, which deliver best classification:
\begin{Code}
min_error = list(error = 1, par = NULL)
for (par in list(par_set_1, ... , par_set_k))
{
  error = ddalpha.getErrorRateCV(data, numchunks = <nc>,
    separator = par$sep, depth = par$depth,
    other_par = par$other_par )
  if(error < min_error$error)
    min_error = list(error = error, par = par)
}
\end{Code}

\section*{Acknowledgments}

The authors want to thank Karl Mosler for his valuable suggestions that have substantially improved the present work.
The authors would like to express their gratitude to the Cologne Graduate School of Management, Economics and Social Sciences who supported the work of Oleksii Pokotylo and to the Lebesgue Centre of Mathematics who supported the work of Pavlo Mozharovskyi (program PIA-ANR-11-LABX-0020-01).


\input{lit}
\end{document}

%% file: table_alpha.tex
\begin{figure}
\centering

\def\ww{.105\textwnew}
\setlength\tabcolsep{1mm}
\renewcommand{\trimval}{59 73 29 50} 
\begin{tabular}{m{0.55cm}|m{\ww}m{\ww}m{\ww}m{\ww}m{\ww}m{\ww}m{\ww}m{\ww}}
  & $y$ & $x^2$& $xy$ & $y^2$ & $x^3$ & $x^2y$ & $xy^2$ & $y^3$ \\
  \hline
$x$ \vfill
& \begin{overpic}[tics = 10, keepaspectratio=true,width = \ww, Trim=\trimval, clip, page = 3]{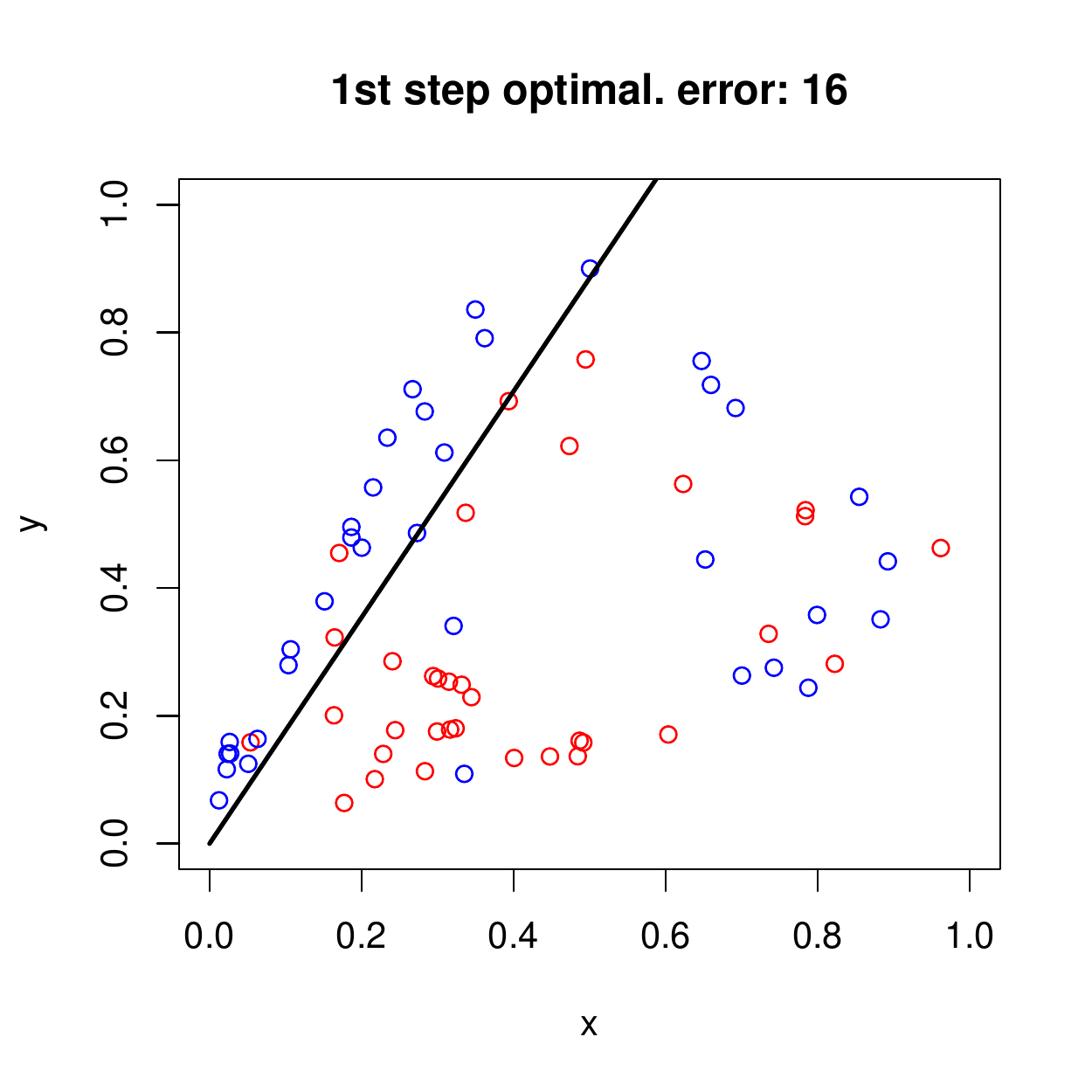} \put (74,65) {16}\end{overpic}
& \begin{overpic}[tics = 10, keepaspectratio=true,width = \ww, Trim=\trimval, clip, page = 4]{ddaplha_steps.pdf} \put (74,65) {26}\end{overpic}
& \begin{overpic}[tics = 10, keepaspectratio=true,width = \ww, Trim=\trimval, clip, page = 5]{ddaplha_steps.pdf} \put (74,65) {23}\end{overpic}
& \begin{overpic}[tics = 10, keepaspectratio=true,width = \ww, Trim=\trimval, clip, page = 6]{ddaplha_steps.pdf} \put (74,65) {19}\end{overpic}
& \begin{overpic}[tics = 10, keepaspectratio=true,width = \ww, Trim=\trimval, clip, page = 7]{ddaplha_steps.pdf} \put (74,65) {26}\end{overpic}
& \begin{overpic}[tics = 10, keepaspectratio=true,width = \ww, Trim=\trimval, clip, page = 8]{ddaplha_steps.pdf} \put (74,65) {23}\end{overpic}
& \begin{overpic}[tics = 10, keepaspectratio=true,width = \ww, Trim=\trimval, clip, page = 9]{ddaplha_steps.pdf} \put (74,65) {23}\end{overpic}
& \begin{overpic}[tics = 10, keepaspectratio=true,width = \ww, Trim=\trimval, clip, page = 10]{ddaplha_steps.pdf} \put (74,65) {21}\end{overpic} \\
$y$ \vfill
&
& \begin{overpic}[tics = 10, keepaspectratio=true,width = \ww, Trim=\trimval, clip, page = 11]{ddaplha_steps.pdf} \put (74,65) {18}\end{overpic}
& \begin{overpic}[tics = 10, keepaspectratio=true,width = \ww, Trim=\trimval, clip, page = 12]{ddaplha_steps.pdf} \put (74,65) {26}\end{overpic}
& \begin{overpic}[tics = 10, keepaspectratio=true,width = \ww, Trim=\trimval, clip, page = 13]{ddaplha_steps.pdf} \put (74,65) {23}\end{overpic}
& \begin{overpic}[tics = 10, keepaspectratio=true,width = \ww, Trim=\trimval, clip, page = 14]{ddaplha_steps.pdf} \put (74,65) {19}\end{overpic}
& \begin{overpic}[tics = 10, keepaspectratio=true,width = \ww, Trim=\trimval, clip, page = 15]{ddaplha_steps.pdf} \put (74,65) {26}\end{overpic}
& \begin{overpic}[tics = 10, keepaspectratio=true,width = \ww, Trim=\trimval, clip, page = 16]{ddaplha_steps.pdf} \put (74,65) {23}\end{overpic}
& \begin{overpic}[tics = 10, keepaspectratio=true,width = \ww, Trim=\trimval, clip, page = 17]{ddaplha_steps.pdf} \put (74,65) {23}\end{overpic} \\
$x^2$ \vfill
&
&
& \begin{overpic}[tics = 10, keepaspectratio=true,width = \ww, Trim=\trimval, clip, page = 18]{ddaplha_steps.pdf} \put (74,65) {16}\end{overpic}
& \begin{overpic}[tics = 10, keepaspectratio=true,width = \ww, Trim=\trimval, clip, page = 19]{ddaplha_steps.pdf} \put (74,65) {16}\end{overpic}
& \begin{overpic}[tics = 10, keepaspectratio=true,width = \ww, Trim=\trimval, clip, page = 20]{ddaplha_steps.pdf} \put (74,65) {26}\end{overpic}
& \begin{overpic}[tics = 10, keepaspectratio=true,width = \ww, Trim=\trimval, clip, page = 21]{ddaplha_steps.pdf} \put (74,65) {23}\end{overpic}
& \begin{overpic}[tics = 10, keepaspectratio=true,width = \ww, Trim=\trimval, clip, page = 22]{ddaplha_steps.pdf} \put (74,65) {19}\end{overpic}
& \begin{overpic}[tics = 10, keepaspectratio=true,width = \ww, Trim=\trimval, clip, page = 23]{ddaplha_steps.pdf}  \put (74,65) {17}\end{overpic} \\
$xy$ \vfill
&
&
&
& \begin{overpic}[tics = 10, keepaspectratio=true,width = \ww, Trim=\trimval, clip, page = 24]{ddaplha_steps.pdf} \put (74,65) {16}\end{overpic}
& \begin{overpic}[tics = 10, keepaspectratio=true,width = \ww, Trim=\trimval, clip, page = 25]{ddaplha_steps.pdf} \put (74,65) {18}\end{overpic}
& \begin{overpic}[tics = 10, keepaspectratio=true,width = \ww, Trim=\trimval, clip, page = 26]{ddaplha_steps.pdf} \put (74,65) {26}\end{overpic}
& \begin{overpic}[tics = 10, keepaspectratio=true,width = \ww, Trim=\trimval, clip, page = 27]{ddaplha_steps.pdf} \put (74,65) {23}\end{overpic}
& \begin{overpic}[tics = 10, keepaspectratio=true,width = \ww, Trim=\trimval, clip, page = 28]{ddaplha_steps.pdf}  \put (74,65) {19}\end{overpic} \\
$y^2$ \vfill
&
&
&
&
& \begin{overpic}[tics = 10, keepaspectratio=true,width = \ww, Trim=\trimval, clip, page = 29]{ddaplha_steps.pdf} \put (74,65) {17}\end{overpic}
& \begin{overpic}[tics = 10, keepaspectratio=true,width = \ww, Trim=\trimval, clip, page = 30]{ddaplha_steps.pdf} \put (74,65) {18}\end{overpic}
& \begin{overpic}[tics = 10, keepaspectratio=true,width = \ww, Trim=\trimval, clip, page = 31]{ddaplha_steps.pdf} \put (74,65) {26}\end{overpic}
& \begin{overpic}[tics = 10, keepaspectratio=true,width = \ww, Trim=\trimval, clip, page = 32]{ddaplha_steps.pdf}  \put (74,65) {23}\end{overpic} \\
$x^3$ \vfill
&
&
&
&
&
& \begin{overpic}[tics = 10, keepaspectratio=true,width = \ww, Trim=\trimval, clip, page = 33]{ddaplha_steps.pdf} \put (74,65) {16}\end{overpic}
& \begin{overpic}[tics = 10, keepaspectratio=true,width = \ww, Trim=\trimval, clip, page = 34]{ddaplha_steps.pdf} \put (74,65) {16}\end{overpic}
& \begin{overpic}[tics = 10, keepaspectratio=true,width = \ww, Trim=\trimval, clip, page = 35]{ddaplha_steps.pdf}  \put (74,65) {16}\end{overpic} \\
$x^2y$ \vfill
&
&
&
&
&
&
& \begin{overpic}[tics = 10, keepaspectratio=true,width = \ww, Trim=\trimval, clip, page = 36]{ddaplha_steps.pdf} \put (74,65) {16}\end{overpic}
& \begin{overpic}[tics = 10, keepaspectratio=true,width = \ww, Trim=\trimval, clip, page = 37]{ddaplha_steps.pdf}  \put (74,65) {16}\end{overpic} \\
$xy^2$ \vfill
&
&
&
&
&
&
&
& \begin{overpic}[tics = 10, keepaspectratio=true,width = \ww, Trim=\trimval, clip, page = 38]{ddaplha_steps.pdf}  \put (74,65) {16}\end{overpic} \\
\hline
$F_1$ \vfill
&
& \begin{overpic}[tics = 10, keepaspectratio=true,width = \ww, Trim=\trimval, clip, page = 39]{ddaplha_steps.pdf} \put (74,65) {15}\end{overpic}
& \begin{overpic}[tics = 10, keepaspectratio=true,width = \ww, Trim=\trimval, clip, page = 40]{ddaplha_steps.pdf} \put (74,65) {16}\end{overpic}
& \begin{overpic}[tics = 10, keepaspectratio=true,width = \ww, Trim=\trimval, clip, page = 41]{ddaplha_steps.pdf} \put (74,65) {16}\end{overpic}
& \begin{overpic}[tics = 10, keepaspectratio=true,width = \ww, Trim=\trimval, clip, page = 42]{ddaplha_steps.pdf} \put (74,65) {13}\end{overpic}
& \begin{overpic}[tics = 10, keepaspectratio=true,width = \ww, Trim=\trimval, clip, page = 43]{ddaplha_steps.pdf} \put (74,65) {15}\end{overpic}
& \begin{overpic}[tics = 10, keepaspectratio=true,width = \ww, Trim=\trimval, clip, page = 44]{ddaplha_steps.pdf} \put (74,65) {16}\end{overpic}
& \begin{overpic}[tics = 10, keepaspectratio=true,width = \ww, Trim=\trimval, clip, page = 45]{ddaplha_steps.pdf}  \put (74,65) {16}\end{overpic} \\

\end{tabular}

\caption{The steps of the $\alpha$-procedure. The number of errors is shown in the right top corner of each plot.
Here we denote the depth of a point w.r.t.~red and blue classes by $x$ and $y$, respectively.
The two-dimensional spaces are shown for each pair of properties. On the first step all pairs of properties are considered, on the second step the remaining features are taken together with the first feature $F_1$. In this example properties $x$ and $y$ are selected on the first step and $x^3$ on the second.}

\label{tab:gt}
\end{figure}

%% file: tables_custom.tex
\def\hsp{-0.8em}

\begin{table}[h!]
\begin{center}
    {\small
\begin{tabular}{|ll|}
\hline
\code{.<name>_validate}&\multicolumn{1}{|l|}{}\\\cline{1-1}
  \multicolumn{2}{|l|}{validates parameters passed to \code{ddalpha.train} and passes them to the \code{ddalpha} object.}\\
    &\\[\hsp]
    IN:&\\
    \code{ddalpha} & {the ddalpha object, containing the data and settings}\\
    \code{<custom parameters>} &  {parameters that are passed to the user-defined method}\\
    \code{...} & {other parameters (mandatory)}\\
    &\\[\hsp]
    OUT:&\\
    \code{list()} & {list of output parameters, after the validation is finished }\\&{these parameters are stored in the \code{ddalpha} object}\\
\hline
  \code{.<name>_learn}&\multicolumn{1}{|l|}{}\\\cline{1-1}
  \multicolumn{2}{|l|}{trains the depth}\\
    &\\[\hsp]
    IN:&\\
    \code{ddalpha} & {the ddalpha object containing the data and settings}\\
    &\\[\hsp]
    MODIFIES:&\\
    \code{ddalpha} & store the calculated statistics in the \code{ddalpha} object \\
    depths & calculate the depths of each pattern, e.g. \\
        & \code{for (i in 1:ddalpha$numPatterns) }\\
        & \code{\ \ ddalpha$patterns[[i]]$depths = }\\
        & \code{\ \ \ \ .<name>_depths(ddalpha, }\\
        & \code{\ \ \ \ \ \ ddalpha$patterns[[i]]$points)}\\
    &\\[\hsp]
    OUT:&\\
    \code{ddalpha} & {the updated \code{ddalpha} object}\\

\hline

  \code{.<name>_depths}&\multicolumn{1}{|l|}{}\\\cline{1-1}
  \multicolumn{2}{|l|}{calculates the depths}\\
    &\\[\hsp]
    IN:&\\
    \code{ddalpha} & {the ddalpha object containing the data and settings} \\
    \code{objects} & {the objects for which the depths are calculated} \\
    &\\[\hsp]
    OUT:&\\
    \code{depths} & {the calculated depths for each object (rows),}\\&{ with respect to each class (columns)}\\

\hline
  Usage: & \code{ddalpha.train(data, depth = "<name>",}\\
         & \code{\ \ <custom parameters>, ...)}\\
\hline
\end{tabular}}
  \caption{Definition of a custom depth function}\label{tab:custom_depth}
  \end{center}
\end{table}

\begin{table}[h!]
  \begin{center}
    {\small
\begin{tabular}{|ll|}
\hline
  \code{.<name>_validate}&\multicolumn{1}{|l|}{}\\\cline{1-1}
  \multicolumn{2}{|l|}{validates parameters passed to \code{ddalpha.train} and passes them to the \code{ddalpha} object}\\
    &\\[\hsp]
    IN:&\\
    \code{ddalpha} & {the ddalpha object containing the data and settings}\\
    \code{<custom parameters>} &  {parameters that are passed to the user-defined method}\\
    \code{...} & {other parameters (mandatory)}\\
    &\\[\hsp]
    OUT:&\\
    \code{list()} & {list of output parameters, after the validation is finished,}\\&{ these parameters are stored in the \code{ddalpha} object.}\\
    &\\
    &\code{methodSeparatorBinary = F} \\
    &in case of a multiclass classifier\\

\hline

  \code{.<name>_learn}&\multicolumn{1}{|l|}{}\\\cline{1-1}
  \multicolumn{2}{|l|}{trains the classifier. Is different for binary and multiclass classifiers.}\\
    &\\[\hsp]
    IN:&\\
    \code{ddalpha} & {the ddalpha object, containing the data and settings}\\
    \code{index1} & {(only for binary) index of the first class}\\
    \code{index2} & {(only for binary) index of the second class}\\
    \code{depths1} & {(only for binary) depths of the first class w.r.t. all classes}\\
    \code{depths2} & {(only for binary) depths of the second class w.r.t. all classes}\\
        &\\[\hsp]
        & depths w.r.t. only given classes are received by \\&\code{    depths1[,c(index1, index2)]}\\
        &\\[\hsp]
        & for multiclass separator the depths are accessible via \\&\code{  ddalpha$patterns[[i]]$depths}\\
        &\\[\hsp]
    OUT:&\\
    \code{classifier} & { the trained \code{classifier} object}\\

\hline

  \code{.<name>_classify}&\multicolumn{1}{|l|}{}\\\cline{1-1}
  \multicolumn{2}{|l|}{classifies the objects}\\
    &\\[\hsp]
    IN:&\\
    \code{ddalpha} & {the ddalpha object, containing the data and global settings} \\
    \code{classifier} & {the previously trained classifier}\\
    \code{objects} & {the objects (depths) that are classified} \\
    &\\[\hsp]
    OUT:&\\
      \code{result} & {a vector with classification results:}\\
        & {positive values for class \code{"classifier$index1"} (binary) or} \\
        & {the indices of a pattern in \code{ddalpha} (multiclass)}\\

\hline
Usage: &\\
binary & \code{ddalpha.train(data, separator = "<name>",}\\ & \code{\ \ aggregation.method = <any>,}\\ & \code{\ \ <custom parameters>, ...)}\\
multiclass & \code{ddalpha.train(data, separator = "<name>",}\\ & \code{\ \ aggregation.method = "none",}\\ & \code{\ \ <custom parameters>, ...)}\\
\hline
\end{tabular}}
  \caption{Definition of a custom separator}\label{tab:custom_separator}
  \end{center}
\end{table}